\documentclass[a4paper,11pt]{article}
\pdfoutput=1 

\usepackage{jheppub} 

\usepackage[T1]{fontenc} 
\newcommand{\mee}{\langle m_{ee}\rangle}
\usepackage{amstext,amssymb}
\usepackage{amsmath}
\usepackage{graphicx}
\usepackage{xspace}
\usepackage{color}
\usepackage{units}
\usepackage{slashed} 

\title{\boldmath Neutrino Mass and Neutrinoless double beta decay in SO(10) GUT with Pati-Salam symmetry}
\author[a]{M. Sruthilaya}
\author[a]{, Rukmani Mohanta}
\author[b]{, Sudhanwa Patra}

\affiliation[a]{School of Physics,  University of Hyderabad, Hyderabad - 500046,  India}
\affiliation[b]{Center of Excellence in Theoretical and Mathematical Sciences, \\
Siksha `O' Anusandhan University, Bhubaneswar-751030, India}

\emailAdd{msruthi28@gmail.com}
\emailAdd{rmsp@uohyd.ernet.in}
\emailAdd{sudha.astro@gmail.com}

\abstract{
We demonstrate how a class of non-supersymmetric $SO(10)$ GUT with asymmetric left-right theory 
$SU(2)_L \times U(1)_R \times U(1)_{B-L} \times SU(3)_C$ and Pati-Salam theory $SU(2)_L \times 
SU(2)_R \times SU(4)_C$ as intermediate symmetry breaking steps leads to successful gauge coupling 
unification satisfying proton decay constraints. The motivation behind this work is two fold: 
firstly to study the renormalization group evolution equations for gauge couplings by keeping right-handed 
neutral gauge boson $Z_R$ around LHC energy range leading interesting dilepton searches at collider 
while fixing charge partner of the gauge boson $W_R$ at very high scale; secondly to explain 
neutrino masses and associated lepton number violating process like neutrinoless double beta 
decay in three possible cases depending on how $SU(2)_L \times U(1)_R \times U(1)_{B-L} \times 
SU(3)_C$ breaks down to SM. The presence of Pati-Salam symmetry and Pati-Salam symmetry with D-parity 
(discrete left-right symmetry leading to $g_L=g_R$) at highest scale is to allow two gauge couplings  
and thereby ensuring precision unification for gauge couplings. We focus 
on neutrino mass and neutrinoless double beta decay for one particular case where TeV scale asymmetric 
left-right theory is spontaneously broken down to SM with non-zero VEV of both Higgs doublets with 
$B-L=-1$ and Higgs triplets with $B-L=2$. We include one extra fermion singlet per generation in order 
to implement gauged extended seesaw where light neutrino mass is governed by natural type-II seesaw 
mechanism whereas type-I seesaw contribution is exactly canceled out. Since light neutrino mass formula 
is independent of Dirac neutrino mass matrix, the value of Dirac neutrino mass is taken to be up-type quark mass matrix which is 
a characteristics of Pati-Salam symmetry relating quarks with leptons. This large value of Dirac neutrino 
mass can contribute to neutrinoless double beta decay, non-unitarity effects in leptonic sector and 
lepton flavor violation. We present analytic relation for effective Majorana mass parameter and corresponding 
half-life arising from new physics contributions due to purely left-handed currents through exchange of 
heavy right-handed neutrinos and sterile neutrinos. We numerically estimate effective Majorana mass parameter 
and half-life vs. lightest neutrino mass and derive lower bound on lightest neutrino mass by saturating 
with experimental bounds like GERDA Phase-II, KamLANDZen and EXO.
}

\begin{document} 
\maketitle
\flushbottom
\newpage
\section{Introduction}
The Standard Model (SM) of particle physics, a well developed theory whose predictions are verified with high precision, seems to be the low energy 
regime of a more fundamental theory. There are some observations like tiny neutrino mass, existence of dark matter, baryon asymmetry etc., urge for 
physics beyond the standard model. In the SM  all particles get their mass through Higgs mechanism and neutrinos are strictly massless due to the absence 
of right handed partner, a must entity to get mass through Higgs mechanism. Dark matter is a matter whose existence is known only through gravitational 
effect and none of the standard model particles can accommodate the observed dark matter density. Also the CP violation in the SM is insufficient to explain 
the matter-antimatter asymmetry of the Universe. All these  indicate that particle content of the standard model has to be extended.

The neutrino mass can be explained if right-handed neutrinos are added to the SM. Since right handed neutrinos are singlets under SM 
symmetry group, they can have Majorana mass which violate $B-L$, an accidental symmetry of the SM by two units. A very high Majorana mass 
of right-handed neutrinos give rise to tiny mass of light neutrinos through Type-I  seesaw mechanism~\cite{Minkowski:1977sc, 
Mohapatra:1979ia, Yanagida:1979as, GellMann:1980vs,Schechter:1980gr} with not so small Yukawa coupling between Higgs and neutrinos. But in this case Majorana mass is an arbitrary parameter 
unlike Dirac mass which is originated by spontaneous breaking of electroweak symmetry. Majorana mass of right handed neutrinos can be generated 
exactly in the same way as mass of other particles if the SM  gauge symmetry is extended by an $SU(2)_R$ symmetry and SM particle content 
by a Higgs which is a triplet under the new symmetry. Other than Type-I seesaw, there are many seesaw mechanisms to attain tiny active neutrino mass 
and all demand additional new particles to the SM particle content. Type-II seesaw~\cite{Magg:1980ut, Schechter:1980gr, Cheng:1980qt, Lazarides:1980nt, Mohapatra:1980yp,Wetterich:1981bx, Antusch:2004xy} 
is one among them, which requires existence of three Higgs fields belonging to the triplet representation of $SU(2)_L$.

 Left-right symmetric~\cite{Pati:1974yy,Mohapatra:1974gc, Senjanovic:1975rk, Deshpande:1990ip} models are examples of popular models 
 where SM symmetry is extended by $SU(2)_R~$ symmetry and right-handed neutrinos get Majorana mass when neutral component 
 of right-handed Higgs triplets get vacuum expectation value. Both these models reveal parity is conserved at very high energies and 
 is spontaneously broken at somewhere above the electroweak scale \cite{Senjanovic:1975rk, Mohapatra:1980qe}, hence give the origin 
 of parity violation in the standard model. The left-right symmetric model--when spontaneous symmetry breaking occurs at few TeV--offers 
 numerous weak interaction phenomenology with TeV scale spectrum of extra gauge bosons $W^\pm_R$, $Z_R$, right-handed neutrinos and 
 associated scalars. Keeping charged gauge boson $W_R$ and neutral gauge boson $Z_R$ within LHC limit, TeV scale left-right symmetry 
 provides  testable consequences for collider signals in the gauge sector~\cite{Keung:1983uu,  Ferrari:2000sp,  Schmaltz:2010xr, Nemevsek:2011hz, 
 Chen:2011hc, Chakrabortty:2012pp, Das:2012ii, AguilarSaavedra:2012gf, Han:2012vk, Chen:2013fna, Rizzo:2014xma,  Deppisch:2014zta, 
 Deppisch:2015qwa, Gluza:2015goa, Ng:2015hba, Patra:2015bga,  Dobrescu:2015qna, Brehmer:2015cia, Dev:2015pga,  Coloma:2015una, Deppisch:2015cua,  
 Dev:2015kca, Mondal:2015zba, Aguilar-Saavedra:2015iew,    Lindner:2016lpp, Lindner:2016lxq, Mitra:2016kov, Anamiati:2016uxp,  
 Khachatryan:2014dka, Aad:2015xaa, Khachatryan:2016jqo} as well as Higgs sector~\cite{Gunion:1989in, Deshpande:1990ip, Polak:1991vf, 
 Barenboim:2001vu, Azuelos:2004mwa, Jung:2008pz, Bambhaniya:2013wza, Dutta:2014dba, Bambhaniya:2014cia,   Bambhaniya:2015wna, 
 Dev:2016dja, ATLAS:2014kca, CMS:2016cpz, ATLAS:2016pbt,Hati:2017aez,Borah:2010zq,Kuchimanchi:2017bfm,Dev:2016vle}. Many attempts have been made in the context of left-right symmetry 
 with and without spontaneous D-parity breaking to give new physics contribtions to neutrinoless double beta decay ($0\nu\beta\beta$)~\cite{Mohapatra:1980yp, 
 Mohapatra:1981pm, Picciotto:1982qe, Hirsch:1996qw, Arnold:2010tu, Tello:2010am,  Chakrabortty:2012mh, Nemevsek:2012iq, Patra:2012ur, Awasthi:2013ff, 
 Barry:2013xxa, Dev:2013vxa, Huang:2013kma, Dev:2014xea, Ge:2015yqa, Borah:2015ufa,  Awasthi:2015ota,   Horoi:2015gdv, Bambhaniya:2015ipg, 
 Gu:2015uek, Borah:2016iqd, Awasthi:2016kbk,Borah:2017inr,Deppisch:2017vne,Parida:2012sq,Patra:2010ks,
Borah:2009ra,Patra:2009wc,Borgohain:2017inp,Ahmed:2017pqa}, low-energy charged lepton flavor violation (cLFV)~\cite{Riazuddin:1981hz, Pal:1983bf, Mohapatra:1992uu, 
 Cirigliano:2004mv, Cirigliano:2004tc, Bajc:2009ft, Tello:2010am, Das:2012ii, Barry:2013xxa, Dev:2013oxa,  Borah:2013lva, Chakrabortty:2015zpm, 
 Awasthi:2015ota, Bambhaniya:2015ipg, Borah:2016iqd, Lindner:2016bgg, Bonilla:2016fqd} and electric dipole moment (EDM)~\cite{Ecker:1983dj, 
 Nieves:1986uk, Frere:1991jt, Nemevsek:2012iq, Dev:2014xea, Maiezza:2014ala}. A few attempts have also made in left-right symmetric model to accommodate 
 stable component of dark matter~\cite{Heeck:2015qra,Garcia-Cely:2015quu,Patra:2015qny,Borah:2016uoi,Borah:2016ees,Borah:2017xgm,Patra:2015vmp,Borah:2017xgm,Borah:2016uoi,
Deppisch:2016scs,Bandyopadhyay:2017uwc,Arbelaez:2017ptu}.

 Models based on Left-Right symmetry are also successful in explaining 
 the existence of dark matter and baryon asymmetry of  the Universe. The recent model discussed in Ref.~\cite{Volovik:2003kh} explains both baryon asymmetry 
 accounting for matter-antimatter asymmetry of the universe and existence of dark matter based on Pati-Salam symmetry with four generations of fermions. 
 In this model, fourth generation neutrino forms dark matter and its number density is shown to account for the major part of dark matter in the Universe.  
 In \cite{Gu:2010yf} baryon asymmetry and existence of dark matter is explained by a model based on left-right symmetry and  \cite{Heeck:2015qra} shows TeV scale dark matter can be accommodated 
 in the left-right symmetric model. Even though these models solve many problems of the standard model they do not explain why there are 
 interactions with different strength and particles with a wide range of mass, a grand unified theory (GUT) is an answer to these questions. 
 Grand Unified Theory (GUT) tells at very high energy  all the three interactions of SM have same strength and  because of different renormalization group evolution 
 with energy due to the particle content of the model,  end up with different strengths at present.

Standard model is a very successful theory at low energy and hence, any GUT candidate  should have SM gauge group as its sub group so that 
SM will be retained at electroweak scale. Since the rank of SM gauge group is four, the rank of gauge group of GUT should be greater than 
or equal to four. Hence the smallest group which can be the gauge group of GUT is $SU(5)$ \cite{Babu:1992ia}. $SO(10)$, $E_6$ etc. are other 
candidates for GUT gauge group. Along with explaining the experimental observations there are some predictions of GUT such as proton decay. 
Non super-symmetric $SU(5)$ GUT is ruled out as it predicts proton lifetime much below the experimental value whereas both SO(10) and $E_6$ 
predict correct proton life time in their non-supersymmetric version.  Among many models, $SO(10)$ GUT is a promising theory as it unifies all the fermion 
content of SM along with right-handed neutrino in a 16 dimensional spinorial representation, thereby giving a common origin for their mass. 
 There are a number of ways in which SO(10) can be broken to SM, which justify experimental observations and are unique by their predictions.  
 
 A GUT with Pati-Salam or Left-Right symmetries as intermediate symmetry will be of great interest. 
In models based on left-right symmetry or Pati-Salam symmetry, discrete parity symmetry breaks along 
with $SU(2)_R$ symmetry and embedding of these models in SO(10) GUT requires parity breaking scale 
to be very high to get the observed value of $\sin^2\theta_W$ \cite{Georgi:1974sy}, hence any effect 
due to the right handed current will be highly suppressed which makes the theory un-testable. 
Introduction of D-parity, symmetry between left and right chiral fields in $SO(10)$ makes parity 
braking scale to decouple from  that of $SU(2)_R$ symmetry \cite{Rizzo:1981dm,Chang:1983fu,Chang:1984uy,
Chang:1984qr}. In such models D-parity breaks at very high energy and $SU(2)_R$ at much lower 
energy, giving rise to a testable effect of right handed currents at energies accessible to current experiments.  

In the present work, we consider a non-supersymmetric $SO(10)$ GUT with intermediate symmetry groups 
such as Pati-Salam symmetry $SU(2)_L \times SU(2)_R \times SU(4)_C$ and TeV scale asymmetric left-right 
theory $SU(2)_L \times U(1)_R \times U(1)_{B-L} \times SU(3)_C$ manifesting in an extra
right-handed neutral gauge boson $Z_R$ which might be detected by ongoing search experiments at the
Large Hadron Collider. Depending on spontaneous symmetry breaking of $SU(2)_L \times U(1)_R 
\times U(1)_{B-L} \times SU(3)_C$ down to SM, we study three different scenarios with particular 
choice of Higgs and comment on neutrino mass and lepton number violating processes like neutrinoless 
double beta decay accordingly. We focus on gauged extended seesaw mechanism for the case where asymmetric 
left-right symmetry breaks down to SM via both Higgs doublet plus triplet and the fermion sector comprises 
of usual quarks and leptons plus one sterile fermions per generation. The mass formula for light neutrinos 
is governed by type-II seesaw dominance while type-I seesaw is exactly canceled out in the diagonalization 
method. We plan to discuss neutrino mass and neutrinoless double beta decay within type-II seesaw dominance 
relating light and heavy mass eigenvalues assuming Dirac neutrino mass matrix equals to up-type quark mass 
matrix. We wish to derive bound on lightest neutrino mass from new physics contributions to neutrinoless 
double beta decay arising from exchange of right-handed neutrinos and sterile neutrinos. 
 
The plan of the paper is organized as following way. In Sec-II we briefly discuss the symmetry breaking 
chain of $SO(10)$ with asymmetric left-right model and Pati-Salam symmetry as intermediate steps of breaking 
and present the one-loop RGEs for gauge couplings along with their matching conditions at different symmetry 
breaking scales and important mass relations. We discuss possible ways to implement the symmetry breaking of 
$U(1)_R\times U(1)_{B-L}\rightarrow U(1)_Y$, gauge coupling unification, neutrino mass and other low energy 
phenomenon in Sec-III, Sec-IV and Sec-V respectively. In Sec-III and Sec-IV, we examine gauge coupling 
unification and commented on neutrino mass and neutrinoless double beta decay. In Sec-V, we study 
successful gauge coupling unification with intermediate mass scales such that they generate neutrino 
mass via type-II seesaw mechanism. We emphasize this case with discussion on neutrinoless double decay and derive lower bounds on lightest neutrino mass by saturating experimental bounds. Towards end, we summarize 
our results and conclude in Sec-VI.

\section{SO(10) GUT with Pati-Salam symmetry}
The idea here is to discuss neutrino mass and associated lepton number violating process like 
neutrinoless double beta decay ($0\nu \beta \beta$) in a non-supersymmetric $SO(10)$ GUT. The popular symmetry breaking 
is $SO(10) \to \mathcal{G}_{SM}$ without having any intermediate symmetry breaking steps. There 
are also other symmetry breaking chain $SO(10) \to \mathcal{G}_{I} \to \mathcal{G}_{SM}$ where the 
intermediate symmetry breaking could be a three gauge groups theory like Pati-Salam 
theory $SU(2)_L \times SU(2)_R \times SU(4)_C$  or four gauge groups theory like left-right 
theory $SU(2)_L \times SU(2)_R \times U(1)_{B-L} \times SU(3)_C$. Instead of manifest left-right 
symmetry group $SU(2)_L \times SU(2)_R \times U(1)_{B-L} \times SU(3)_C$, one may consider 
asymmetric left-right theory with gauge group $SU(2)_L \times U(1)_R \times U(1)_{B-L} \times SU(3)_C$ 
as possible intermediate symmetry breaking step where $SU(2)_R$ has broken down to $U(1)_R$ without 
breaking rank of the gauge group. One such symmetry breaking chain for $SO(10)$ GUT having  asymmetric 
left-right theory as well as Pati-Salam symmetry as possible subgroups is given by
\begin{eqnarray}
\label{eq:BreakingChain}
&&SO(10)\,   \mathop{\longrightarrow}^{M_U} \mathcal{G}_{2_L 2_R 4_C D}\, 
             \mathop{\longrightarrow}^{M_{D_P}} \mathcal{G}_{2_L 2_R 4_C}\, 
             \mathop{\longrightarrow}^{M_{W_R}} \mathcal{G}_{2_L 1_R 1_{B-L} 3_C}\, 
  	     \mathop{\longrightarrow}^{M_{Z_R}} \mathcal{G}_{2_L 1_Y 3_C}\, \big(\mbox{SM}\big)
  	     \mathop{\longrightarrow}^{M_{W}} \mathcal{G}_{1_{Q} 3_C}, \hspace*{1.0 truecm}
\end{eqnarray}
where we have defined
\begin{eqnarray}
&&\mathcal{G}_{2_L 2_R 4_C D}\equiv SU(2)_L \times SU(2)_R \times SU(4)_C \times D, \nonumber \\
&&\mathcal{G}_{2_L 2_R 4_C}\equiv SU(2)_L \times SU(2)_R \times SU(4)_C,\nonumber \\
&&\mathcal{G}_{2_L 1_R 1_{B-L} 3_C}\equiv SU(2)_L \times U(1)_R \times U(1)_{B-L} \times SU(3)_C, \nonumber \\
&&\mathcal{G}_{2_L 1_{Y} 3_C}\equiv SU(2)_L \times U(1)_Y \times SU(3)_C.\nonumber 
\end{eqnarray}

In this work, we consider how the left-right asymmetric gauge group $SU(2)_L \times U(1)_R \times U(1)_{B-L} \times SU(3)_C$ 
breaks down to the Standard Model gauge group $SU(2)_L \times U(1)_Y \times SU(3)_C$ with the choice of Higgs field. 
We found that there are three possible ways to implement the symmetry breaking i.e., $U(1)_R \times U(1)_{B-L} \to 
U(1)_Y$ via
\begin{itemize}
\item Higgs doublet $H^0_{R} (1,1/2,-1,1)$ with $B-L$ charge $-1$
\item Higgs triplet $\Delta^0_R (1,-1,2,1)$ with $B-L$ charge $2$
\item combination of Higgs doublet $H_R$ as well as Higgs triplet $\Delta_R$.
\end{itemize}
\subsection{RG evolution for the gauge couplings}
We study the RGEs for the gauge couplings in a non-supersymmetric SO(10) GUT with 
$\mathcal{G}_{2_L 1_Y 3_C}$, $\mathcal{G}_{2_L 1_R 1_{B-L} 3_C}$, $\mathcal{G}_{2_L 2_R 4_C}$ and 
$\mathcal{G}_{2_L 2_R 4_C D}$ as intermediate symmetry breaking steps where the evolution equations 
for running coupling constants at one-loop level is given by
\begin{equation}{\label{4.1}}
\mu\,\frac{\partial g_{i}}{\partial \mu}=\frac{a_i}{16 \pi^2} g^{3}_{i},
\end{equation}
which can be written in the form
\begin{equation}{\label{4.2}}
\frac{1}{\alpha_{i}(\mu_{2})}=\frac{1}{\alpha_{i}(\mu_{1})}-\frac{a_{i}}{2\pi} \ln \left( \frac{\mu_2}{\mu_1}\right),
\end{equation}
where we denote $\alpha_{i}=g_{i}^{2}/4\pi$ as the fine structure constant for
$i$-th gauge group and $\mu_1, \mu_2$ are two different energy scales with
$\mu_2 > \mu_1$. The master formula for one-loop beta-coefficients $a_i$ determining the evolution
of gauge couplings at one-loop order is given as
\begin{eqnarray}{\label{4.3}}
	&&a_i= - \frac{11}{3} \mathcal{C}_{2}(G) 
				 + \frac{2}{3} \,\sum_{R_f} T(R_f) \prod_{j \neq i} d_j(R_f) 
  + \frac{1}{3} \sum_{R_s} T(R_s) \prod_{j \neq i} d_j(R_s)\;.
\label{oneloop_bi}
\end{eqnarray}
In the above mentioned formula for $a_i$, $\mathcal{C}_2(G)$ represents the  quadratic Casimir operator for the
gauge bosons in their adjoint representation with values
\begin{equation}{\label{4.4}}
	\mathcal{C}_2(G) \equiv \left\{
	\begin{matrix}
		N & \text{if } SU(N), \\
    0 & \text{if }  U(1)\;.
	\end{matrix}\right.
\end{equation}
Similarly, $T(R_f)$ and $T(R_s)$ are defined as the Dynkin indices of the
irreducible representation for $R_{f,s}$ for a given fermion and scalar,
respectively, 
\begin{equation}{\label{4.5}}
	T(R_{f,s}) \equiv \left\{
	\begin{matrix}
		1/2 & ~~~~~~\text{if } R_{f,s} \text{ is fundamental}, \\
    N   & \text{if } R_{f,s} \text{ is adjoint}, \\
		0   & \text{if } R_{f,s} \text{ is singlet}\;.
	\end{matrix}\right.
\end{equation}
Here $d(R_{f,s})$ stands for the dimension of a given representation $R_{f,s}$
under all other $SU(N)$ gauge groups except the $i$-th~gauge group.

\subsection{Matching conditions for inverse fine structure constants:}
The inverse fine structure constants satisfy the following matching conditions at different scales. At the scale
 $\mu =M_{Z_R}$,  $U(1)_R\times U(1)_{B-L}$ broken down to $U(1)_Y$ while $SU(3)_C$ and $SU(2)_L$ remain the same. Hence the matching conditions are given as
\begin{eqnarray}\label{eq:4}
[\alpha_{3C}^{-1}(M_{Z_R})]_{{\cal G}_{SM}} &=&  [\alpha_{3C}^{-1}(M_{Z_R})]_{{\cal G}_{2113}},~~~~
{[\alpha_{2L}^{-1}(M_{Z_R})]}_{{\cal G}_{SM}} = [\alpha_{2L}^{-1}(M_{Z_R})]_{{\cal G}_{2113}}\nonumber\\
{[\alpha_{Y}^{-1}(M_{Z_R})]}_{{\cal G}_{SM}} &=& \left [\frac{3}{5}\alpha_{1R}^{-1}(M_{Z_R})+\frac{2}{5}\alpha_{B-L}^{-1}(M_{Z_R})\right ]_{{\cal G}_{2113}}
\end{eqnarray}
Similarly at $\mu = M_{W_R}$, $SU(4)_C$ breaks to $SU(3)\times U(1)_{B-L}$ and $SU(2)_R$ breaks to $U(1)_R$ keeping $SU(2)_R$ intact. Hence, at this energy structure constants of $SU(3)_C$ and $U(1)_{B-L}$ are same and equal to that of $SU(4)_C$ while structure constants of $U(1)_R$ and $SU(2)_L$ of ${\cal G}_{2113}$ shares the same value with that of $SU(2)_R$ and $SU(2)_L$ of ${\cal G}_{224}$ respectively, i.e.,  
\begin{eqnarray}\label{eq:5}
[\alpha_{3C}^{-1}(M_{W_R})]_{{\cal G}_{2113}}& =& [\alpha_{4C}^{-1}(M_{W_R})]_{{\cal G}_{224}}\;,~~~~
{[\alpha_{2L}^{-1}(M_{W_R})]}_{{\cal G}_{2113}}= [\alpha_{2L}^{-1}(M_{W_R})]_{{\cal G}_{224}}\;,\nonumber\\
{[\alpha_{1R}^{-1}(M_{W_R})]}_{{\cal G}_{2113}}&=&[\alpha_{2R}^{-1}(M_{W_R})]_{{\cal G}_{224}}\;,~~~~
{[\alpha_{B-L}^{-1}(M_{W_R})]}_{{\cal G}_{2113}}=[\alpha_{4c}^{-1}(M_{W_R})]_{{\cal G}_{224}}\;.\hspace{0.5truecm}
\end{eqnarray}
From $\mu=M_{D_P}$ onwards D-parity is respected hence, the structure constants of $SU(2)_R$ and $SU(2)_L$ share same value for all energies above D-parity breaking scale $M_{D_P}$. Hence the matching conditions of structure constants at $M_{D_P}$ are 
\begin{eqnarray}\label{eq:6}
[\alpha_{4C}^{-1}(M_{D_P})]_{{\cal G}_{224}}& =& [\alpha_{4C}^{-1}(M_{D_P})]_{{\cal G}_{224D}},~~~~
{[\alpha_{2L}^{-1}(M_{D_P})]}_{{\cal G}_{224}}= [\alpha_{2L}^{-1}(M_{D_P})]_{{\cal G}_{224D}}\nonumber\\
{[\alpha_{2R}^{-1}(M_{D_P})]}_{{\cal G}_{224}}&=&{[\alpha_{2L}^{-1}(M_{D_P})]}_{{\cal G}_{224D}}
=[\alpha_{2R}^{-1}(M_{D_P})]_{{\cal G}_{224D}}\;.
\end{eqnarray}
At $\mu=M_U$ all the structure constants converge to a single one, the structure constant of SO(10), hence the matching conditions satisfy
\begin{eqnarray}\label{eq:7}
{[\alpha_{2L}^{-1}(M_{U})]}_{{\cal G}_{224D}}
=[\alpha_{2R}^{-1}(M_{U})]_{{\cal G}_{224D}} =  
{[\alpha_{4C}^{-1}(M_U)]}_{{\cal G}_{224D}}=[\alpha_{10}^{-1}(M_U)]_{SO(10)}\;.
\end{eqnarray}
 We denote here the one loop beta coefficients in the mass range $M_W-M_{Z_R}$ as $a_i=\{a_{2L}, a_{Y}, a_{3C}\}$, from $M_{Z_R}-M_{W_R}$ 
as $a^\prime_i=\{a^\prime_{2L}, a^\prime_{1R}, a^\prime_{BL}, a^\prime_{3C}\}$, from $M_{W_R}-M_{D_P}$ as $a^{\prime \prime}_i=\{a^{\prime \prime}_{2L}, a^{\prime \prime}_{2R}, a^{\prime \prime}_{4C}\}$ and 
from $M_{D_P}-M_{U}$ as $a^{\prime \prime \prime}_i=\{a^{\prime \prime \prime}_{2L}, a^{\prime \prime \prime}_{2L}, a^{\prime \prime \prime}_{4C}\}$. 
Solving Eq. (\ref{4.1}) along with the matching conditions (\ref{eq:4}), (\ref{eq:5}), (\ref{eq:6}) and (\ref{eq:7}) gives the standard model gauge coupling constants as
\begin{eqnarray}\label{eq:8}
\alpha_{3C}^{-1}(M_W)&=& [\alpha_{10}^{-1}(M_U)]_{SO(10)}+ \frac{a_{3C}}{2 \pi} \ln \left (\frac{M_{Z_R}}{M_W} \right )
+ \frac{a_{3C}^{\prime}}{2 \pi} \ln \left (\frac{M_{W_R}}{M_{Z_R}} \right )\nonumber\\
&+&   \frac{a_{4C}^{ \prime \prime}}{2 \pi} \ln \left (\frac{M_{D_P}}{M_{W_R}} \right )
+ \frac{a_{4C}^{\prime \prime \prime  }}{2 \pi} \ln \left (\frac{M_U}{M_{D_P}} \right ).
\end{eqnarray}
\begin{eqnarray}\label{eq:9}
\alpha_{2L}^{-1}(M_W)&=& [\alpha_{10}^{-1}(M_U)]_{SO(10)}+ \frac{a_{2L}}{2 \pi} \ln \left (\frac{M_{Z_R}}{M_W} \right )
+ \frac{a_{2L}^{\prime}}{2 \pi} \ln \left (\frac{M_{W_R}}{M_{Z_R}} \right )\nonumber\\
&+&  \frac{a_{2L}^{\prime \prime }}{2 \pi} \ln \left (\frac{M_{D_P}}{M_{W_R}} \right )
+ \frac{a_{2L}^{\prime \prime \prime }}{2 \pi} \ln \left (\frac{M_U}{M_{D_P}} \right ).
\end{eqnarray}
\begin{eqnarray}\label{eq:10}
\alpha_{Y}^{-1}(M_W)&=& [\alpha_{10}^{-1}(M_U)]_{SO(10)}+ \frac{a_Y}{2 \pi} \ln \left (\frac{M_{Z_R}}{M_W} \right )
+ \frac{1}{2 \pi} \left (\frac{3}{5} 
a_{1R}^{\prime }+\frac{2}{5} a_{BL}^{ \prime } \right )
\ln \left (\frac{M_{W_R}}{M_{Z_R}} \right ) \nonumber\\
&+&  \frac{1}{2 \pi} \left (\frac{3}{5} 
a_{2R}^{  \prime \prime }+\frac{2}{5} a_{4C}^{ \prime \prime } \right )
\ln \left (\frac{M_{D_P}}{M_{W_R}} \right ) 
+ \frac{1}{2 \pi} \left (\frac{3}{5} 
a_{2L}^{ \prime \prime \prime }+\frac{2}{5} a_{4C}^{\prime \prime \prime } \right )
\ln \left (\frac{M_U}{M_{D_P}} \right ).\hspace*{0.8truecm} 
\end{eqnarray}

\subsection{Solutions for $M_{D_P}$ and $M_U$ by fixing $M_{Z_R}$ around TeV scale}
At the energy scale $M_W$, the gauge coupling constants satisfy the following relations
\begin{eqnarray}
\frac{1}{\alpha_{em}}\left (\sin^2 \theta_W - \frac{3}{8} \right )
= \frac{5}{8} \left (\frac{1}{\alpha_{2L}(M_W)}- \frac{1}{\alpha_{Y}(M_W)} \right ),\label{eq:11}
\end{eqnarray}
and
\begin{eqnarray}
8\left (\alpha_s^{-1} - \frac{3}{8} \alpha_{em}^{-1} \right ) = 8 \alpha_{3C}^{-1} -3 \alpha_{2L}^{-1} -5 \alpha_Y^{-1}.\label{eq:12}
\end{eqnarray}
With the above conditions along with Eqs. (\ref{eq:8}) to (\ref{eq:10}) we obtain
\begin{equation}
{\cal A}_D \ln \left (\frac{M_{D_P}}{M_W} \right ) + {\cal A}_U \ln \left ( \frac{M_U}{M_W} \right ) =D_0 \label{eq:13}
\end{equation}
where the different parameters in Eq. (\ref{eq:13}) are given as
\begin{equation}\label{eq:14}
D_0=\frac{16 \pi}{\alpha_{em}} \left ( \sin^2 \theta_W - \frac{3}{8} \right )-{\cal A}_{Z_R} \ln \left ( \frac{M_{Z_R}}{M_W} \right )
-{\cal A}_{W_R} \ln \left ( \frac{M_{W_R}}{M_W} \right ),
\end{equation}
and
\begin{eqnarray}\label{eq:15}
{\cal A}_{Z_R} &=& (5 a_{2L} - 5 a_Y) -(5 a_{2L}^{\prime}- 3 a_{1R}^{\prime} - 2 a_{B-L}^{\prime}),\nonumber\\
{\cal A}_{W_R} &=& (5 a_{2L}^{\prime} - 3a_{1R}^{\prime} -2 a_{B-L}^{\prime} ) -
(5 a_{2L}^{\prime \prime}- 3 a_{2R}^{\prime \prime} - 2 a_{4C}^{\prime \prime}),\nonumber\\
{\cal A}_{D} &= &(5 a_{2L}^{\prime \prime}- 3 a_{2R}^{\prime \prime} - 2 a_{4 C}^{\prime  \prime})
-(2 a_{2L}^{\prime \prime \prime }- 2 a_{4 C}^{\prime \prime \prime}),\nonumber\\
{\cal A}_U &= &(2 a_{2L}^{\prime  \prime \prime}- 2 a_{4 C}^{\prime \prime \prime}).
\end{eqnarray}
Analogously, one can also obtain the relation
\begin{equation}\label{eq:16}
{\cal B}_D \ln \left (\frac{M_{D_P}}{M_W} \right ) + {\cal B}_U \ln \left ( \frac{M_U}{M_W} \right ) =D_1,
\end{equation}
where
\begin{equation}\label{eq:17}
D_1=16 \pi \left ( \alpha_s^{-1} - \frac{3}{8} \alpha_{em}^{-1} \right )
-{\cal B}_{Z_R} \ln \left ( \frac{M_{Z_R}}{M_W} \right )
-{\cal B}_{W_R} \ln \left ( \frac{M_{W_R}}{M_W} \right ),
\end{equation}
and
\begin{eqnarray}\label{eq:18}
{\cal B}_{Z_R} &=& (8a_{3C} - 3a_{2L} - 5 a_Y) -(8 a_{3C}^{\prime} -3 a_{2L}^{\prime}- 3 a_{2R}^{\prime} - 2 a_{B-L}^{\prime}),\nonumber\\
{\cal B}_{W_R} &=& (8 a_{3C}^{\prime} -3 a_{2L}^{\prime} - 3a_{1R}^{\prime} -2 a_{B-L}^{\prime} ) -
(6 a_{4C}^{\prime \prime} -3 a_{2L}^{\prime \prime}- 3 a_{2R}^{\prime \prime} ),\nonumber\\
{\cal B}_D &= &(6 a_{4 C}^{\prime \prime } -3 a_{2L}^{\prime  \prime}- 3 a_{2R}^{\prime  \prime}  )
-(6 a_{4 C}^{\prime  \prime  \prime} - 6 a_{2L}^{ \prime \prime \prime } ),\nonumber\\
{\cal B}_U &= &(6 a_{4 C}^{\prime  \prime \prime} -6 a_{2L}^{\prime \prime \prime} )\;.
\end{eqnarray}
Thus with Eqs. (\ref{eq:13}) and (\ref{eq:16}) we get the relations
\begin{eqnarray}
\ln \left(\frac{M_U}{M_W} \right ) &=& \frac{D_1 {\cal A}_D -D_0 {\cal B}_D}{{\cal B}_U {\cal A}_D -{\cal A}_U {\cal B}_D}\;,\nonumber\\
\ln \left(\frac{M_D}{M_W} \right ) &=& \frac{D_0 {\cal B}_U -D_1 {\cal A}_U}{{\cal B}_U {\cal A}_D -{\cal A}_U {\cal B}_D}\;.
\end{eqnarray}

\section{Grand Unification with intermediate $U(1)_R \times U(1)_{B-L} \to U(1)_Y$ breaking via Higgs doublets $H_{L,R}$}
The important stage of intermediate symmetry breaking $U(1)_R \times U(1)_{B-L} \to U(1)_Y$ is done via Higgs doublets $H_{L,R}$ with $B-L$ charge $-1$ 
at a scale of right-handed neutral gauge boson mass $M_{Z_R}$. We embed this intermediate left-right 
symmetry group $SU(2)_L \times U(1)_R \times U(1)_{B-L} \times SU(3)_C$ into a $SO(10)$-GUT with Pati-Salam 
intermediate symmetry which can be best understood via the following way,
\begin{eqnarray*}
&\hspace*{-3.0cm} SO(10) &\nonumber \\
&\hspace*{-0.5cm} \downarrow \hspace*{-0.0cm} \langle \eta(1,1,1) \rangle \subset{54_H} &\nonumber\\
&
SU(2)_{L}^{}\times SU(2)_{R}^{} \times SU(4)_C^{} \times D \quad (g_L = g_R) &\nonumber\\
&\hspace*{-0.5cm} \downarrow \hspace*{-0.0cm} \langle \sigma(1,1,1) \rangle \subset{210_H} &\nonumber\\
&
SU(2)_{L}^{}\times SU(2)_{R}^{} \times SU(4)_C^{}\quad (g_L \neq g_R) &\nonumber\\
&
\hspace*{-0.2cm} \downarrow \hspace*{-0.0cm} \langle \Sigma(1,3,15) \rangle \subset{210_H} &\nonumber\\
&
 SU(2)_{L}^{}\times U(1)_{R}^{}\times U(1)_{B-L}^{} \times SU(3)_C^{}\quad (g_L \neq g_R) &\nonumber\\
&
\hspace*{0.5cm} \downarrow   \langle H^0_R(1,1/2,-1,1) \rangle \subset{16_H} &\nonumber\\
&
\hspace*{-1.8cm} SU(2)_{L}^{}\times U(1)_{Y}^{} \times SU(3)_C^{} \nonumber\\
&
\hspace*{-0.6cm}\downarrow\langle \phi(2,\frac{1}{2},1) \rangle \subset{10_H} &\nonumber\\
&\hspace*{-2.8cm} U(1)_{Q}^{}\times SU(3)_C^{} \,.&
\end{eqnarray*}
At first, the GUT scale symmetry breaking $SO(10) \to \mathcal{G}_{2_L 2_R 4_C D}$ is achieved by giving non-zero vacuum expectation value to D-parity even singlet $\eta (1,1,1) \subset 54_H$. 
As a result, the Higgs sector is symmetric under left-right invariance from D-parity scale ($M_{D_P}$) to unification scale ($M_U$) leading to $g_{L} = g_{R}$. The second step of symmetry breaking 
$\mathcal{G}_{2_L 2_R 4_C D} \to \mathcal{G}_{2_L 2_R 4_C}$ is done by assigning a non-zero VEV to D-parity odd singlet $\sigma(1,1,1) \subset 210_H$ which is part of $SO(10)$ GUT \cite{Chang:1983fu,Chang:1984uy}. 
The next stage of symmetry breaking $ \mathcal{G}_{2_L 2_R 4_C}  \to \mathcal{G}_{2_L 1_R 1_{B-L} 3_C}$ is occurred by the neutral component of right-handed Higgs $\Sigma_R (1,3,15) \subset 210_H$. 
At this stage of symmetry breaking, $SU(2)_R$ breaks down to $U(1)_R$ without breaking the rank of the gauge group and the $W_R$ gets it mass around Pati-Salam symmetry breaking scale i.e, at 
$M_{W_R}$. The important stage of symmetry breaking is happened by giving non-zero VEV to neutral component of right-handed Higgs doublet $H_R(1,1/2,-1,1) \subset (1,2,\overline{4}) \subset 16_H$. At this stage, 
the extra right-handed neutral gauge boson $Z_R$ gets its mass around few TeV. The last step of symmetry 
breaking is done with SM Higgs doublet $\phi(2,1/2,1) \subset 16_H$ leading to fermions masses and mixing. 

\subsection{Gauge coupling unification and different mass scales}
\begin{figure}[h!]
\begin{center}
\includegraphics[width=0.7\linewidth]{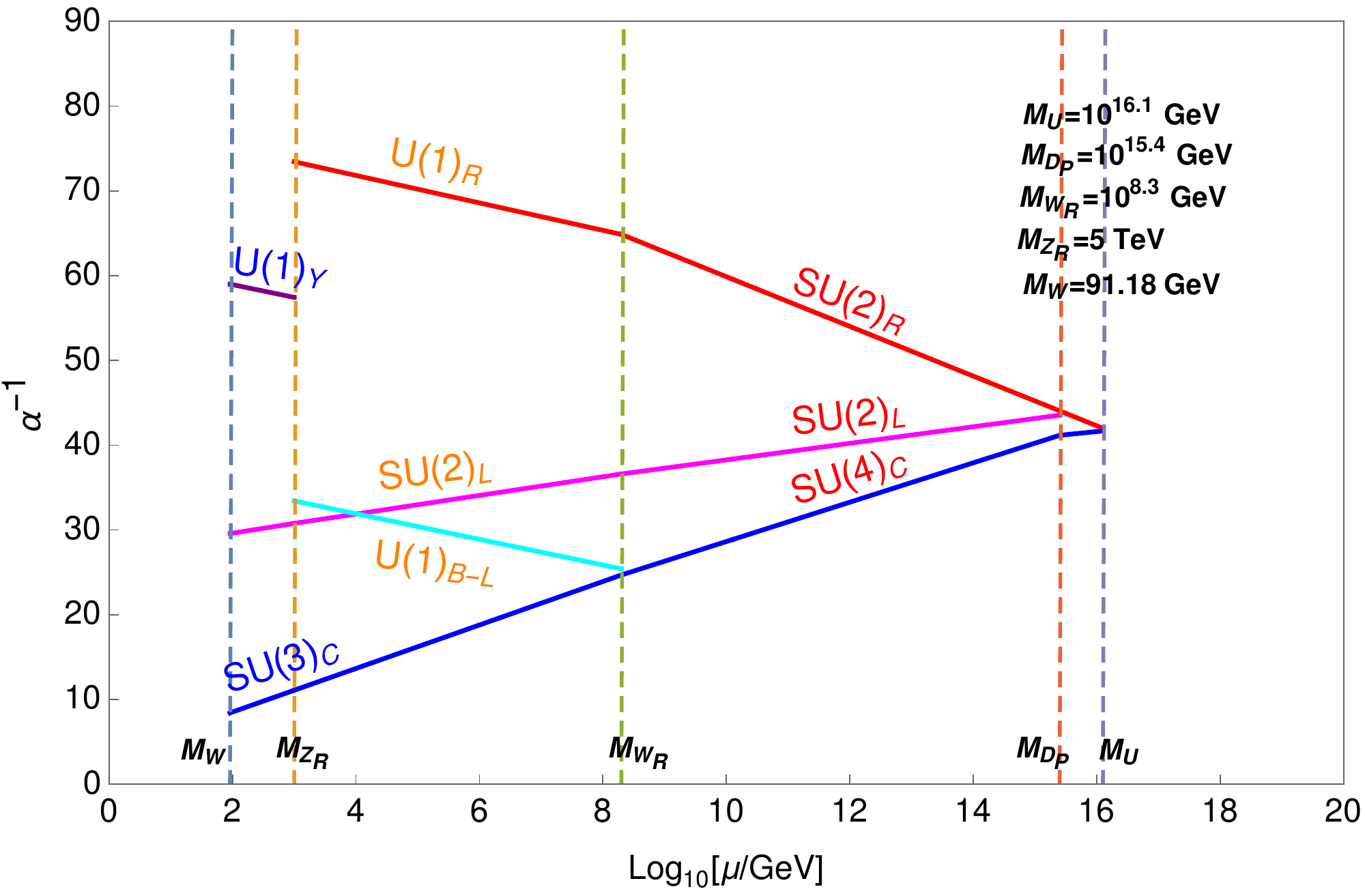}
\caption{  Evolution of inverse fine structure constants $\alpha^{-1}_i$ with variation of energy scale. Gauge coupling running demonstrating successful gauge unification
  within SO(10) GUT with intermediate asymmetric left-right gauge group and Pati-Salam symmetry breaking as intermediate symmetry 
  breaking steps. We fix the right-handed neutral gauge boson scale around 5~TeV and Pati-Salam breaking scale around $10^{8.3}~$GeV while other mass scales like D-parity 
  breaking scale and unification scale are predicted to be $M_{D_P} \approx 10^{15.4}~$GeV and  $M_U \approx10^{16.1}~$GeV. } 
 \label{fig1}
\end{center}
\end{figure}

 We study RGEs for gauge couplings for non-supersymmetric $SO(10)$ GUT with asymmetric left-right theory, Pati-Salam symmetry and Pati-Salam symmetry with D-parity as intermediate 
subgroups and the TeV scale asymmetric left-right theory spontaneously broken down to SM via Higgs doublet with $B-L$ charge $-1$.
The Higgs spectrum and the derived one-loop beta coefficients for this framework are presented in Table.\,\ref{tab1}. The unification plot demonstrating successful gauge coupling unification is displayed 
in Fig. \ref{fig1} with intermediate mass scales as follows,
$$M_{Z_R} \approx \mbox{5\,TeV}, \quad M_{W_R} \approx 10^{8.3}\, \mbox{GeV}, \quad M_{D_P} \approx 10^{15.4}\, \mbox{GeV}, \quad M_{U} \approx 10^{16.1}\, \mbox{GeV}\, .$$
\begin{table}[htb]
\begin{tabular}{|c|c|c|c|}
\hline 
Group & Range of masses &Higgs Content& $a_i$ values\\ 
\hline

${\cal G}_{213}$ & $M_W - M_{Z_R}$ & $\Phi(2,1/2,1)_{10}$ & $a_i= \left (\begin{array}{c}  -19/6 \\ 41/10 \\ -7 \end{array} \right )$ \tabularnewline
\hline

${\cal G}_{2113}$ & $M_{Z_R} - M_{W_R}$ &  $\begin{array}{c} \Phi_1(2,1/2,0,1)_{10},  \Phi_2(2,-1/2,0,1)_{10^\prime}    \\ 
        H_R(1,1/2,-1,1) \end{array}$  & $a_i^{\prime}= \left (\begin{array}{c} -3 \\ 53/12 \\ 33/8 \\ -7 \end{array} \right )$ \tabularnewline
\hline

${\cal G}_{224}$ & $M_{W_R} - M_{D_P}$ & 
 $\begin{array}{c} \Phi_1(2,2,1)_{10},  \Phi_2(2,2,1)_{10^\prime}    \\ 
         H_R(1,2,\overline{4}),  \Sigma_R(1,3,15)_{210} \end{array}$ 
 & $a_i^{\prime \prime}= \left (\begin{array}{c} -8/3 \\  8 (3) \\ -19/3 (-25/3)  \end{array} \right )$ \tabularnewline
\hline

${\cal G}_{224D}$ & $M_{D_P} - M_{U}$ & $\begin{array}{c} 
    \Phi_1(2, 2,1)_{10}, \Phi_2(2,2,1)_{10'} \\
    H_L(2,1,4)_{16} , H_R(1,2,\overline{4})_{16} \\
     \Sigma_L (3,1,15)_{210}, \Sigma_R(1,3,15)_{210}\\ 
     \sigma(1,1,1)_{210}    \end{array}$ & $a_i^{\prime \prime \prime}= \left (\begin{array}{c} 8 (3) \\  8 (3) \\ -2 (-6)
 \end{array} \right )$ \tabularnewline
\hline

\end{tabular}
\caption{Details of the Gauge groups present in the symmetry breaking pattern with corresponding energy ranges, Higgs spectrum and one loop beta coefficients. For the last stage of symmetry breaking to Standard model symmetry is implemented by Higgs doublet $H_R$.  The one loop beta 
           coefficients are derived from Pati-Salam scale onwards by taking complex (real) Higgs representation $(1,3,15)$ where the numbers within parenthesis are 
           for real representation.} 
\label{tab1}
\end{table}

\subsection{Fermion mass fitting}
Since $SO(10)$ grand unified theory unifies matter (15 components of fermions in one generation of SM plus additional right-handed neutrinos) 
in a 16-dimensional spinor and contains left-right symmetry and Pati-Salam symmetry as its subgroups, it provides fermion masses and mixing 
as well as relates quarks and leptons. Here, we intend to discuss the minimal realization of the theory by adding different combination of Higgs 
representation and comment whether or not they can be realistic and predictive. We restrict ourselves to Yukawa interactions of fermions at 
renormalizable level. It is already established that $SO(10)$ GUT with only Higgs $10_H$ is inconsistent with fermion masses and mixing as 
SM doublet contained in $10_H$ is $SU(4)_C$ singlet and thus, results same structure for quarks and leptons. 

With $16_F \otimes 16_F = 10 \oplus 120 \oplus 126$, it is possible to write down three Yukawa interaction terms using $10_H$, $120_H$ and 
$126_H$.   As $10_H$ alone can not explain fermion masses and mixing and within the present scenario with no Higgs triplets contained in $126_H$, 
we are left with the choice $10_H+120_H$. With $10_H$ and $120_H$, the Yukawa interaction Lagrangian is given by
\begin{eqnarray}
\mathcal{L}_{Y} = 16_F \left(Y_{10} \, 10_H + Y_{120} 120_H \right) 16_F\, ,
\end{eqnarray}
with $Y_{10}=Y^T_{10}$ and $Y_{120}=-Y^T_{120}$. The Pati-Salam $SU(2)_L \times SU(2)_R \times SU(4)_C$ decomposition of $10_H$ and $120_H$ 
can be read as
\begin{eqnarray}
&&10_H= (2,2,1) \oplus (1,1,6)\, , \nonumber \\
&&120_H=(2,2,1) \oplus (2,2,15) \oplus (1,1,10) \oplus (1,1,\overline{10}) \oplus (1,3,6) \oplus (3,1,6)\, .
\end{eqnarray}
Using the non-zero VEVs of Higgs field contained in $10_H$ and $120_H$, the fermion masses read as~\cite{Bajc:2005aq},
\begin{subequations}
\begin{eqnarray}
&&M_d= M_0 + M_2\, , \quad \quad M_u = c_0 M_0 + c_2 M_2\;, \\ 
&& M_e=M_0 + c_3 M_2 \,, \quad \quad M^\nu_D = c_0 M_0 + c_4 M_2\;,
\label{eq:mee-std}
\end{eqnarray}
\end{subequations}
where
\begin{subequations}
\begin{eqnarray}
&&M_0 = Y_{10} \langle \left(2,2,1 \right)^d_{10} \rangle\, ,\\
&&M_2 = Y_{120} \left[\langle \left(2,2,1 \right)^d_{120} \rangle+ \langle \left(2,2,15 \right)^d_{120} \rangle \right]\,,  \\
&&c_0 M_0 = Y_{10} \langle \left(2,2,1 \right)^u_{10} \rangle\, ,\\
&&c_2 M_2 = Y_{120} \left[\langle \left(2,2,1 \right)^u_{120} \rangle+ \langle \left(2,2,15 \right)^u_{120} \rangle \right]\,,  \\
&&c_3 = \frac{\langle \left(2,2,1 \right)^d_{120} \rangle -3 \langle \left(2,2,15 \right)^d_{120} \rangle}
                              {\langle \left(2,2,1 \right)^d_{120} \rangle+ \langle \left(2,2,15 \right)^d_{120} \rangle}\, , \\
&&c_4 = \frac{\langle \left(2,2,1 \right)^u_{120} \rangle -3 \langle \left(2,2,15 \right)^u_{120} \rangle}
                              {\langle \left(2,2,1 \right)^d_{120} \rangle+ \langle \left(2,2,15 \right)^d_{120} \rangle}\, .                             
\label{eq:mee-std}
\end{eqnarray}
\end{subequations}
In addition to this Dirac neutrino mass $M^\nu_D$, there could be two loop diagrams which can contribute to  Majorana mass for right-handed 
neutrinos using $10_H$, $16_H$ and $45_V$. The two loop contribution is proportional to $\propto \left(\alpha/\pi \right)^2 Y_{10} \left(\langle16_H\rangle^2/M_{GUT} \right)$ 
which is negligible in our present case with $\langle H_R \rangle \subset 16_H$ around few TeV. Thus, we have only dominant Dirac masses for light neutrinos. 
The sub-eV scale of Dirac neutrino mass can be achieved by suitable adjustment of different Yukawa couplings or fine tuning between $c_0 M_0$ and $c_4 M_2$. 
The details of the numerical fitting for fermion masses and mixing for three generation picture is rather involved and messy which is beyond the scope of 
this paper.

\subsection{Comment on Neutrino Mass and Neutrinoless double beta decay}
The fermion content of the asymmetric left-right model is given by
\begin{eqnarray}
 \ell_{L} &= \begin{pmatrix}\nu_L \\ e_L\end{pmatrix} \, \sim
(\mathbf{2},\mathbf{0},\mathbf{-1}, \mathbf{1})\, , & 
N_R  (\mathbf{1},\mathbf{1/2},\mathbf{-1}, \mathbf{1})\,, \quad  e_R (\mathbf{1},\mathbf{-1/2},\mathbf{-1}, \mathbf{1}) \, , \nonumber  \\[1mm]
 q_{L} &= \begin{pmatrix}u_L \\ d_L\end{pmatrix} \, \sim
(\mathbf{2},\mathbf{0},\mathbf{\frac{1}{3}},\mathbf{3})\, , & 
u_R(\mathbf{1},\mathbf{1/2},\mathbf{1/3}, \mathbf{3}) \, , \nonumber  ~~d_R(\mathbf{1},\mathbf{-1/2},\mathbf{1/3}, \mathbf{3})\;. \nonumber
\end{eqnarray}
The spontaneous symmetry breaking for asymmetric left-right model to SM is done via $\Phi_1(2,1/2,0,1)$, $\Phi_2(2,-1/2,0,1)$ and $H_{R}(1,1/2,-1,1)$. 
No Majorana masses for left-handed as well as right-handed neutrinos are allowed due to absence of Higgs triplets. Thus, we have only Dirac masses for 
charged fermions,
\begin{eqnarray}
&&  M_u =  Y_1 v_1 + Y_2 v^*_2\,, \quad \quad 
     M_d =  Y_1 v_2 + Y_2 v^*_1\,, \quad \quad \nonumber \\
&&  M_e =  Y_3 v_2 + Y_4 v^*_1 \,,\quad \quad
\end{eqnarray}
 as well as for neutrinos
\begin{equation}
M^\nu_D\equiv M_D = Y_3 v_1 + Y_4 v^*_2 \,,
\end{equation}
where, $v_1 (v_2)$ is the vacuum expectation value for $\Phi_1$ ($\Phi_2$) and $Y_i$ are the corresponding Yukawa couplings. 
The sub-eV scale of light active neutrinos is explained by this Dirac neutrino mass by adjusting these couplings and VEVs. 
Since there is no Majorana mass for neutrinos there is no lepton number violation present in the model where symmetry breaking 
of asymmetric left-right models is implemented with Higgs doublets. As a result of this, the lepton number violating process like 
neutrinoless double beta decay is absent.

\section{Grand Unification with intermediate $U(1)_R \times U(1)_{B-L} \to U(1)_Y$ breaking via Higgs triplets $\Delta_{L,R}$}

In this section we discuss the case where breaking  of $U(1)_R\times U(1)_{B-L}$ to $U(1)_Y$ is implemented by nonzero VEV of 
$\Delta_{R}(1,1,-2,1)\subset(1,3,\overline{10})\subset 126_H$, the neutral component of right-handed Higgs triplet. It has a very different 
representation compared to $H_R(1,1/2,-1,1)$ under all symmetry groups present at different stages of symmetry breaking, hence affects 
RG evolution considerably. Moreover $\Delta_R$ belongs to higher dimensional representations of all symmetry groups other than $SU(2)_L$ 
which helps to lower the Pati-Salam and $D$ parity breaking scale down to $10^5$ GeV and $10^{12}$ GeV respectively. A lower $D$ parity 
breaking scale ensures a sizable Type-II seesaw contribution to light neutrino mass. A sizable Type-II seesaw correction to neutrino mass 
makes it independent of quark mass whereas in $SO(10)$ grand unified theories  Dirac neutrino mass and up-type quark mass have same 
origin. Breaking of $SU(2)_R$ symmetry at lower scale allows right handed currents to have detectable effects in low energy phenomenon 
such as neutrinoless double beta decay.    

\subsection{Gauge coupling unification}
\begin{figure}[h!]
\begin{center}
\includegraphics[width=0.7\linewidth]{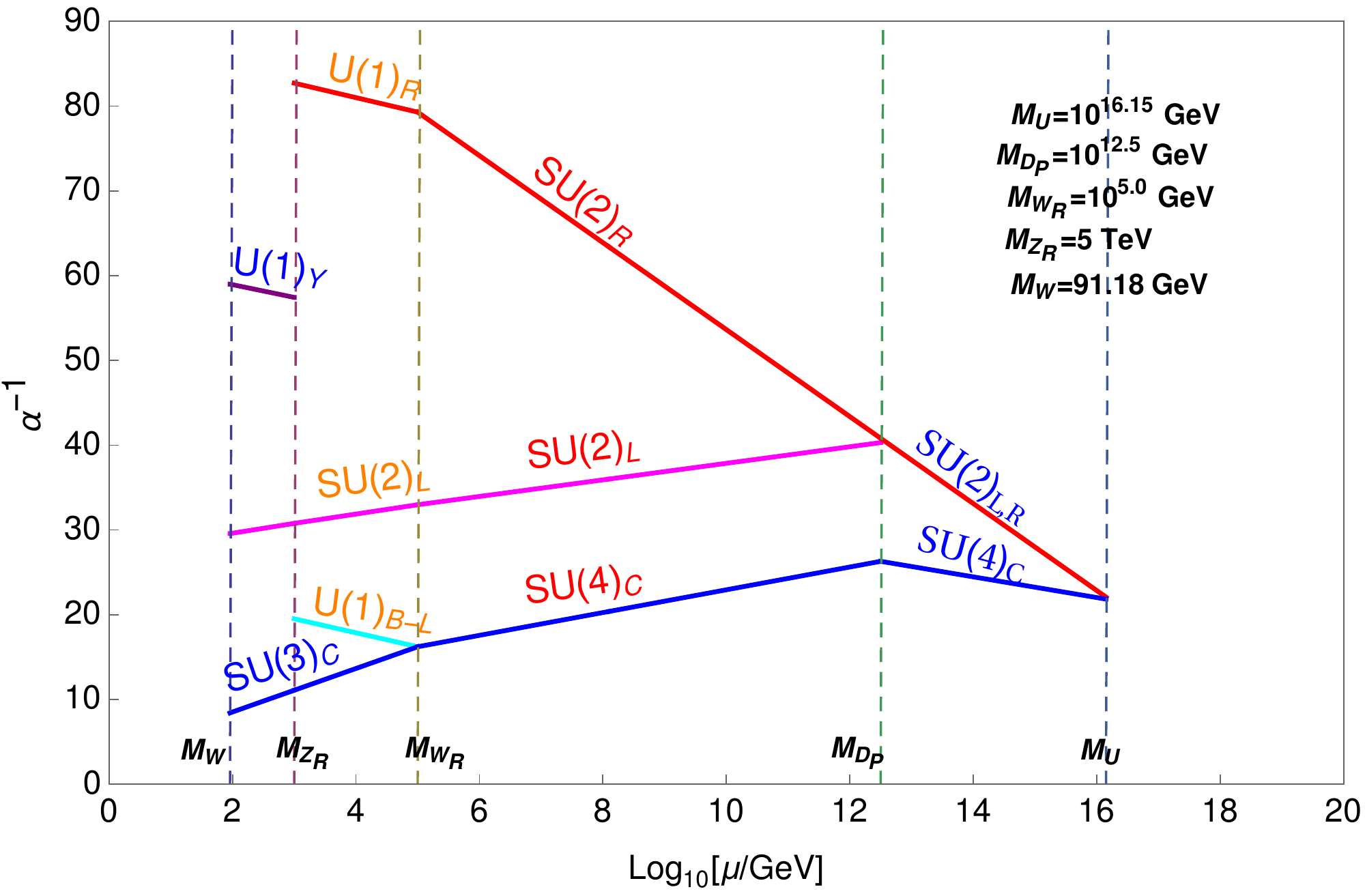}
\caption{ Evolution of inverse fine structure constants $\alpha^{-1}_i$ with variation of energy scale. We fix the right-handed neutral gauge boson scale around 
5~TeV and Pati-Salam breaking scale around $10^{5}~$GeV while other mass scales like D-parity breaking scale and unification scale are predicted by RGEs. 
The D-parity breaking scale $M_{D_P}$ is found to be $10^{12.5}~$GeV leading to induced VEV for scalar triplet  and type-II seesaw contribution to 
light neutrino mass as $m^{II}_\nu = f v_L \approx \mathcal{O}(1) \cdot v^2 \cdot v_R/ 
\left(M^\prime \cdot M_{D_P} \right) \simeq 10^{-3}~$eV while fixing $M^\prime \approx 10^{8}~$GeV. Thus, the type-I contributions to light neutrino mass is dominant than 
the type-II seesaw  contribution having TeV scale of right-handed neutrinos.}  
\label{fig2}
\end{center}
\end{figure}

 We study RGEs for gauge couplings for non-supersymmetric $SO(10)$ GUT with asymmetric left-right theory, Pati-Salam symmetry and Pati-Salam symmetry with D-parity as intermediate 
subgroups and the TeV scale asymmetric left-right theory spontaneously broken down to SM via Higgs triplet with $B-L$ charge $2$.
The Higgs content and one-loop beta coefficients at various stages of symmetry breaking are given in Table.\,\ref{tab2}. The unification plot demonstrating successful gauge coupling unification is displayed 
in Fig. \ref{fig2} with intermediate mass scales as follows,
$$M_{Z_R} \approx \mbox{5\,TeV}, \quad M_{W_R} \approx 10^{5}\, \mbox{GeV}, \quad M_{D_P} \approx 10^{12.5}\, \mbox{GeV}, \quad M_{U} \approx 10^{16.15}\, \mbox{GeV}\, .$$
\begin{table}[htb]
\begin{center}
\begin{tabular}{|c|c|c|c|}
\hline 
~Group~ &~ Range of masses ~ &~ Higgs Content~ & ~~$a_i$ values ~~\tabularnewline
\hline

${\cal G}_{213}$ & $M_W - M_{2R}$ & $\Phi(2,1/2,1)_{10}$ & $a_i= \left (\begin{array}{c}   -19/6 \\ 41/10\\-7 \end{array} \right )$ \tabularnewline
\hline

${\cal G}_{2113}$ & $M_{2R} - M_{W_R}$ & $\begin{array}{c} \Delta_R(1,1,-2,1)_{126}\\ \oplus~ \Phi_1(2,1/2,0,1)_{10} \\
\oplus~ \Phi_2(2,-1/2,0,1)_{10'} \end{array}$ & $a_i^{\prime}= \left (\begin{array}{c}  -3 \\ 14/3 \\ 9/2\\-7 \end{array} \right )$ \tabularnewline
\hline

${\cal G}_{224}$ & $M_{W_R} - M_{D_P}$ & $\begin{array}{c} \Sigma_R(1,3,15)_{210}\\ \oplus~ \Delta_R(1,3,\overline{10})_{126}\\ 
\oplus~ \Phi_1(2, 2,1)_{10}\\ \oplus~ \Phi_2(2,2,1)_{10'} \end{array}$ & $a_i^{\prime \prime}= \left (\begin{array}{c} -8/3 \\ 42/3(9)\\-11/3(-17/3) \end{array} \right )$ \tabularnewline
\hline

${\cal G}_{224D}$ & $M_{D_P} - M_{U}$ & $\begin{array}{c}\eta(1,1,1)_{210} \\ \oplus~ \Sigma_R (1,3,15)_{210}\\ \oplus~ \Sigma_L(3,1,15)_{210}\\ 
\oplus~ \Delta_R(1,3,\overline{10})_{126} \\ \oplus \Delta_L(3,1,10)_{126} \\
\oplus~ \Phi_1(2, 2,1)_{10}\\ \oplus~ \Phi_2(2,2,1)_{10'} \end{array}$ & $a_i^{\prime \prime \prime}= \left (\begin{array}{c}  42/3(9) \\ 42/3(9)\\10/3(-2/3) 
 \end{array} \right )$ \tabularnewline
\hline

\end{tabular}
\caption{ Higgs spectrum with Higgs triplet to break $\mathcal{G}_{2113}$ to $\mathcal{G}_{SM}$ and corresponding one-loop beta coefficients 
              at various energy ranges in the symmetry breaking chain.  Here one loop beta  coefficients are derived from Pati-Salam scale 
              onwards by taking complex (real) Higgs representation $(1,3,15)$ where the numbers within parenthesis are  for real representation.}
              \label{tab2}
\end{center}
\end{table}

\subsection{Fermion mass fitting}
We need $\Delta_R \subset 126_H$, $\Phi_1 \subset 10_H$ and $\Phi_1 \subset 10^\prime_H$ for spontaneous symmetry breaking of asymmetric left-right symmetry breaking. 
The relevant Yukawa Lagrangian at SO(10) level relevant for fermion masses and mixing is given by
\begin{eqnarray}
\mathcal{L}_{Y} = 16_F \left(Y_{10} \, 10_H + Y_{10^\prime} \, 10^\prime_H + Y_{126} 126_H \right) 16_F\, ,
\end{eqnarray}
with $Y_{10}=Y^T_{10}$, $Y_{10^\prime}=Y^T_{10^\prime}$  and $Y_{126}= Y^T_{126}$. The Pati-Salam $SU(2)_L \times SU(2)_R \times SU(4)_C$ decomposition of $10_H$ and $126_H$ 
is given by
\begin{eqnarray}
&&10_H= (2,2,1) \oplus (1,1,6)\, , \nonumber \\
&&10^\prime_H= (2,2,1) \oplus (1,1,6)\, , \nonumber \\
&&126_H= (2,2,15) \oplus (3,1,10) \oplus (1,3,\overline{10}) \oplus (1,1,6)\, .
\end{eqnarray}
With $10_H$, $10^\prime_H$ and $126_H$, the fermion masses are given by
\begin{eqnarray}
&&M_d = v^{10}_d Y_{10} + v^{10^\prime}_d Y_{10^\prime} + v^{126}_d Y_{126}\, , \quad \quad M_u = v^{10}_u Y_{10} + v^{10^\prime}_u Y_{10^\prime}  + v^{126}_u Y_{126} \, , \nonumber \\
&&M_e = v^{10}_d Y_{10} + v^{10^\prime}_d Y_{10^\prime}  - 3 v^{126}_d Y_{126}\, , \quad \quad M^\nu_D = v^{10}_u Y_{10} + v^{10^\prime}_u Y_{10^\prime}  - 3  v^{126}_u Y_{126} \, , \nonumber \\
&&M_L = v_L Y_{126}\, , \nonumber \\
&&M_R = v_R Y_{126}\, .
\end{eqnarray}
We can ignore the type-II seesaw contribution $m^{II}_\nu = f v_L \propto f \beta v^2 v_R /(M' M_{D_P})$ (with large D-parity breaking scale $M_{D_P}$) as compared 
to type-I seesaw contribution $m^{I}_\nu=-M^\nu_D M^{-1}_R M^{\nu^T}_D$. For TeV scale $M_R$, we need small value of $M_D$ which can be obtained 
with fine tuning the Yukawa couplings although the details is beyond the scope of this paper. 

\subsection{Comment on Neutrino Mass}
A pleasant situation arises in which the spontaneous symmetry breaking is done by assigning a non-zero VEV to the neutral 
component of right-handed Higgs triplet $\Delta_R$ carrying $B-L$ charge $2$. The left-right invariance demands the existence 
of left-handed Higgs triplet $\Delta_L$. The Majorana masses for left-handed as well as right-handed neutrinos are generated 
via non-zero VEV of $\Delta_L$ and $\Delta_R$ and thus violate lepton number by two units. We can write down the relevant 
interaction terms connecting scalar triplets and leptons as follows,
\begin{eqnarray}
-\mathcal{L}_{\rm Yuk} &\supset& f_{ij}\left[\overline{(\ell_{Li})^c}\ell_{Lj}\Delta_L+
\overline{(\ell_{Ri})^c}\ell_{Rj}\Delta_R\right]+ \mbox{h.c.}\,.
\label{eqn:LR-Yuk-triplets}
\end{eqnarray}
The Higgs bidoublets will result Dirac neutrino mass matrix connecting both left- and right-handed neutrinos. 
The Majorana mass terms $M_L$ and $M_R$ arise due to presence of scalar triplets. Thus, the neutral lepton 
mass matrix in the basis $\left(\nu_L, N^c_R\right)$ reads as
\begin{equation}
\mathcal{M}_\nu= \left( \begin{array}{cc}
              M_{L} & M_{D}   \\
              M^T_{D} & M_{R}
                      \end{array} \right) \, ,
\label{eqn:numatrix}       
\end{equation}
The resulting light neutrino mass is governed by type-I and type-II seesaw formula as
\begin{eqnarray}
m_{\nu} = -M_{D}\, M_{R}^{-1}\, M_{D}^{T} + M_L = m_{\nu}^I + m_{\nu}^{II}\, ,
\label{type2a}
\end{eqnarray}
where, $m_{\nu}^I$ ($ m_{\nu}^{II}$) is for canonical type-I (type-II) seesaw contribution to light neutrino masses
$$m_{\nu}^I=-M_{D}\, M_{R}^{-1}\, M_{D}^{T}, \quad \quad m_{\nu}^{II}= f\, v_L=f\, \langle \Delta^0_L \rangle\,. $$

\subsection{Comment on Neutrinoless double beta decay}
Due to presence of Majorana nature of neutrinos in the present scenario, we have lepton number violating process like smoking-gun same-sign dilepton signatures at LHC as well as neutrinoless double beta decay 
in low energy experiments  (for details refer to Refs~\cite{Ge:2015bfa,Deppisch:2014zta,Tello:2010am,Deppisch:2014qpa,Patra:2014goa,
Borah:2013lva,Awasthi:2013ff,Patra:2012ur,Chakrabortty:2012mh,Dev:2013vxa,Barry:2013xxa,Nemevsek:2011hz,
Dev:2014iva,Keung:1983uu,Das:2012ii,Bertolini:2014sua,Beall:1981ze,Ge:2015yqa,Deppisch:2015cua,Hirsch:1996qw,Dev:2013oxa,
Dev:2015pga}). In the context of neutrinoless double beta decay, many works consider either type-I seesaw dominance or type-II 
seesaw dominance within manifest left-right symmetric model, in left-right symmetric model with spontaneous D-parity breaking making unequal 
gauge couplings $g_L \neq g_R$. In most of these analysis, the left-right neutrino mixing (active-sterile neutrino mixing) is very much suppressed  
and as a result of this, the cross-section of heavy neutrino production and other lepton number violating processes at LHC are very much suppressed.

\section{Grand Unification with intermediate $U(1)_R \times U(1)_{B-L} \to U(1)_Y$ breaking via Higgs doublets and triplets}
\subsection{Spontaneous symmetry breaking}
Here the intermediate symmetry breaking $U(1)_R\times U(1)_{B-L} \to U(1)_Y$ is implemented with non-zero 
VEVs of both Higgs doublets $H_R \subset 16_H$ and triplets $\Delta_R \subset 126_H$ around few TeV scale. Then we embed this intermediate 
breaking step in a non-supersymmetric $SO(10)$ GUT with Pati-Salam symmetry and Pati-Salam symmetry with D-parity as two other intermediate 
symmetry breaking steps.  The RG evolution equations for gauge coupling constants are almost similar to that in the previous section where 
$U(1)_R\times U(1)_{B-L}$ symmetry is broken by $\Delta_R$ alone. The only difference is the one-loop beta coefficients where both Higgs doublets 
and triplets contribute differently.

With the inclusion of extra sterile singlet neutrinos along with usual left- and right-handed neutrinos, the light neutrino mass formula is governed by 
extended type-II seesaw with the combination of $\Delta_R$ and $H_R$. In this case light neutrinos get mass only through Type-II seesaw even though 
right handed neutrinos get both Dirac and Majorana masses. The canonical type-I seesaw contribution to light neutrino mass is completely canceled out and thereby 
allow large value of Dirac neutrino mass which we will discuss in detail in the subsequent  discussions. Within $SO(10)$ GUT with Pati-Salam intermediate symmetry 
it is not possible to have vanishing Dirac neutrino mass  as Pati-Salam symmetry relates both up-type quark and Dirac neutrino mass matrices with each other. 
Since both Dirac and Majorana masses of right-handed neutrinos are independent of light neutrino mass,  the structure of Dirac and Majorana masses may provide 
the scope of successful Leptogenesis to explain the Baryon Asymmetry of Universe.

\subsection{Gauge coupling unification}
\begin{figure}[htb!]
\begin{center}
\includegraphics[width=0.7\linewidth]{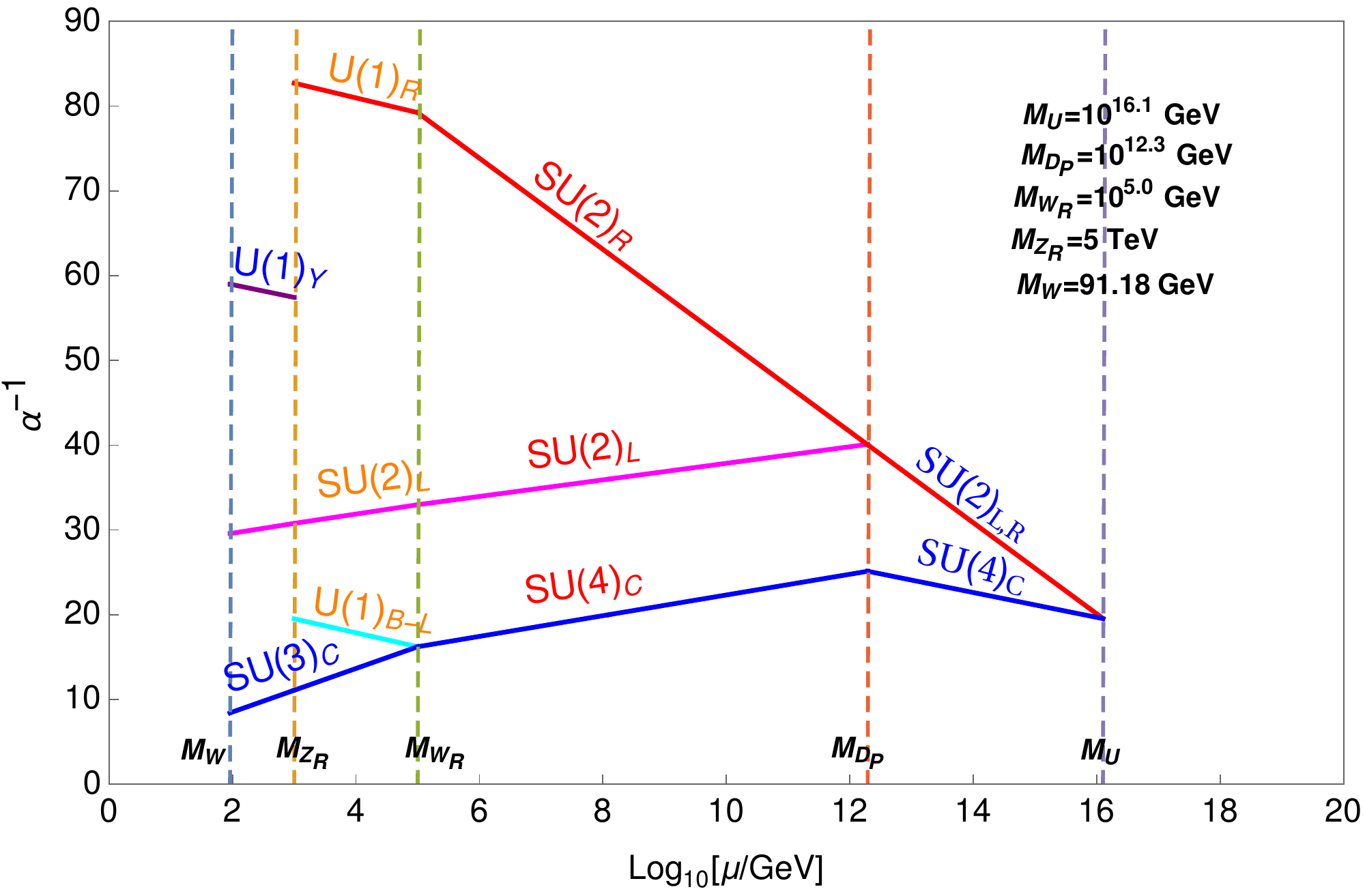}
\caption{  Evolution of inverse fine structure constants $\alpha^{-1}_i$ with variation of energy scale. We fix the right-handed neutral gauge boson scale around 
5~TeV and Pati-Salam breaking scale around $10^{5}~$GeV while other mass scales like D-parity breaking scale and unification scale are predicted by RGEs. 
The D-parity breaking scale $M_{D_P}$ is found to be $10^{12.13}~$GeV leading to induced VEV for scalar triplet  and type-II seesaw contribution to 
light neutrino mass as $m^{II}_\nu = f v_L \approx \mathcal{O}(1) \cdot v^2 \cdot v_R/ 
\left(M^\prime \cdot M_{D_P} \right) \simeq 10^{-3}-10^{-2}~$eV while fixing $M^\prime \approx 10^{8}~$GeV. } 
\label{fig3}
\end{center}
\end{figure}

 We study RGEs for gauge couplings for non-supersymmetric $SO(10)$ GUT with asymmetric left-right theory, Pati-Salam symmetry and Pati-Salam symmetry with D-parity as intermediate 
subgroups and the TeV scale asymmetric left-right theory spontaneously broken down to SM via Higgs triplet with $B-L$ charge $2$.
The Higgs content and one-loop beta coefficients at various stages of symmetry breaking are given in Table.\,\ref{tab3}. The unification plot demonstrating successful gauge coupling unification is displayed 
in Fig. \ref{fig3} with intermediate mass scales as follows,
$$M_{Z_R} \approx \mbox{5\,TeV}, \quad M_{W_R} \approx 10^{5}\, \mbox{GeV}, \quad M_{D_P} \approx 10^{12.3}\, \mbox{GeV}, \quad M_{U} \approx 10^{16.1}\, \mbox{GeV}\, .$$

\begin{table}[htb]
\begin{center}
\begin{tabular}{|c|c|c|c|}
\hline 
~Group~ &~ Range of masses ~ &~Higgs Content~ & ~~$a_i$ values\\
\hline

${\cal G}_{213}$ & $M_W - M_{Z_R}$ & $\Phi (2,1/2,1)_{10}$ & $a_i= \left (\begin{array}{c} -19/6 \\ 41/10\\-7 \end{array} \right )$ \tabularnewline
\hline

${\cal G}_{2113}$ & $M_{Z_R} - M_{W_R}$ & $\begin{array}{c} \Phi_1(2,1/2,0,1)_{10},  \Phi_2(2,-1/2,0,1)_{10^\prime}    \\ 
         \Delta_R(1,1,2,1)_{126} ,    H_R(1,1/2,-1,1) \end{array}$ & $a_i^{\prime}= \left (\begin{array}{c} -3 \\ 57/12 \\ 37/8\\-7 \end{array} \right )$ \tabularnewline
\hline

${\cal G}_{224}$ & $M_{W_R} - M_{D}$ & $\begin{array}{c} 
                         \Phi_1(2, 2,1)_{10},  \Phi_2(2,2,1)_{10^\prime} \\
                         \Delta_R(1,3,\overline{10})_{126},   H_R(1,2,\overline{4})_{16}  \\
                         \Sigma_R(1,3,15)_{210}      \end{array}$ 
                        &
$a_i^{\prime \prime}= \left (\begin{array}{c} -8/3 \\ 44/3(29/3)\\-10/3(-16/3) \end{array} \right )$ \tabularnewline
\hline

${\cal G}_{224D}$ & $M_D - M_{U}$ & $\begin{array}{c}
                         \Phi_1(2, 2,1)_{10},  \Phi_2(2,2,1)_{10^\prime} \\
                         \Delta_R(1,3,\overline{10})_{126},  \Delta_L(3,1,10)_{126}  \\
                         H_R(1,2,\bar{4})_{16}, H_L(2,1,4)_{16} \\
                         \Sigma_R(1,3,15)_{210}, \Sigma_L(3,1,15) _{210} \\
                         \eta(1,1,1)_{210}     \end{array}$  &
$a_i^{\prime \prime \prime}= \left (\begin{array}{c}  44/3(29/3) \\ 44/3(29/3)\\4(0)
 \end{array} \right )$ \tabularnewline
\hline

\end{tabular}
\caption{ Higgs spectrum with Higgs triplet  and doublet to break $\mathcal{G}_{2113}$ to $\mathcal{G}_{SM}$ and corresponding one loop beta coefficients at various 
               energy ranges in the symmetry breaking chain.  The one loop beta 
           coefficients are derived from Pati-Salam scale onwards by taking complex (real) Higgs representation $(1,3,15)$ where the numbers within parenthesis are 
           for real representation.} 
               \label{tab3}
\end{center}
\end{table}

\begin{figure}[htb!]
\begin{center}
\includegraphics[width=0.7\linewidth]{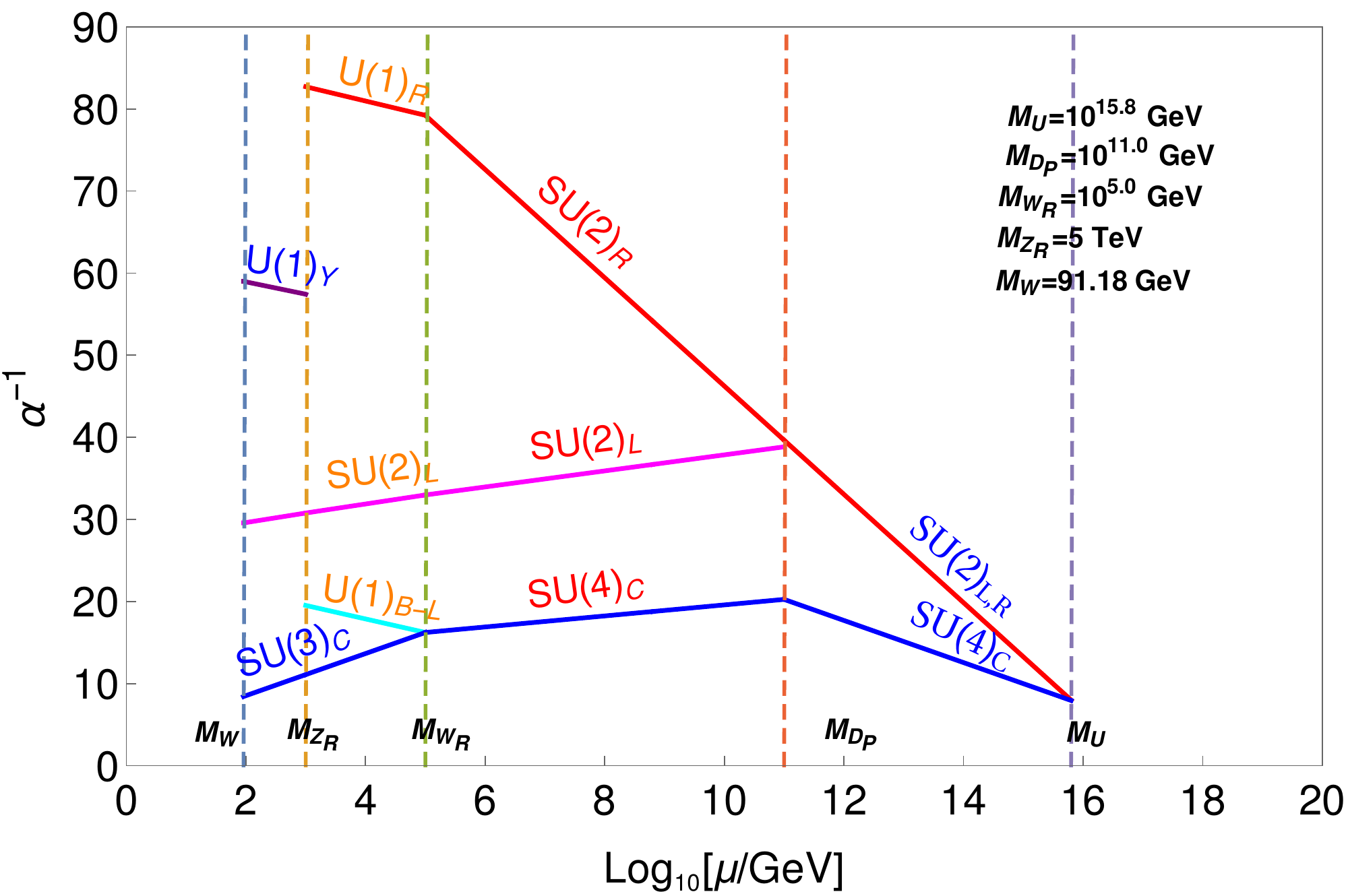}
\caption{ 
 We demonstrate how one can lower the D-parity breaking scale by introducing another scalar triplet contained in $(1,3,\overline{10})$ from Pati-Salam symmetry breaking 
scale onwards which may enhance the scale of induced VEV of left-handed scalar triplet and the type-II seesaw contribution. In the presence if additional fermion singlet per generation, 
the type-I seesaw contribution is exactly canceled out leading to light neutrino masses via natural type-II seesaw dominance with large value of Dirac neutrino mass which gives observable 
effects. } 
\label{fig3a}
\end{center}
\end{figure}

\subsection{Fermion mass fitting}
We need both triplet $\Delta_R \subset 126_H$ as well as doublet $H_R \subset 16_H$ along with two bidoublets $\Phi_1 \subset 10_H$ and $\Phi_1 \subset 10^\prime_H$ 
for spontaneous symmetry breaking of asymmetric left-right symmetry breaking.   Under Pati-Salam symmetry $SU(2)_L \times SU(2)_R \times SU(4)_C$, the quarks and leptons of 
$i$-th generation put in a single representation,
\begin{equation}
F_{L(R)} = \begin{pmatrix} 
                 u_r  & u_g  & u_b  & \nu_e \\
                 d_r  & d_g  & d_b  & e
                  \end{pmatrix}_{L(R)}\, .
\end{equation}
Here $F_L \equiv (2,1,4)$, $F_R \equiv (1,2,\overline{4})$ and $F_L \oplus F_R$ forms $16_F$ spinorial representation of SO(10). 
We know $(2,1,4) \otimes (1,2,\overline{4}) = (2,2,1) \oplus (2,2,15)$, the Dirac masses for quarks and leptons of the form $Y \overline{F_L} \Phi F_R$ 
are generated through non-zero VEVs of $(2,2,1)$ of $10_H$ and $10^\prime_H$, $(2,2,15)$ of $126_H$. Since $(2,2,1) \subset 126_H$ is very small, 
we need two bidoublets contained in $10_H$ and $10^\prime_H$ for correct charged fermion masses and mixing. On the other hand, the non-zero 
VEV of $(3,1,10)$ and $(1,3,\overline{10})$ give Majorana masses for left-handed and right-handed neutrinos.

The  SO(10) Yukawa Lagrangian for fermion masses is given by
\begin{eqnarray}
\mathcal{L}_{Y} = 16_F \left(Y_{10} \, 10_H + Y_{10^\prime} \, 10^\prime_H + Y_{126} 126_H \right) 16_F\, ,
\end{eqnarray}
with $Y_{10}=Y^T_{10}$ , $Y_{10^\prime}=Y^T_{10^\prime}$  and $Y_{126}= Y^T_{126}$. The Pati-Salam $SU(2)_L \times SU(2)_R \times SU(4)_C$ decomposition of $10_H$ and $126_H$ 
is given by
\begin{eqnarray}
&&10_H= (2,2,1) \oplus (1,1,6)\, , \nonumber \\
&&10^\prime_H= (2,2,1) \oplus (1,1,6)\, , \nonumber \\
&&16_H= (2,1,4) \oplus (1,2,\overline{4})\, , \nonumber \\
&&126_H=(2,2,15) \oplus (3,1,10) \oplus (1,3,\overline{10}) \oplus (1,1,6)\, .
\end{eqnarray}

The resulting fermion masses is given by 
\begin{eqnarray}
&&M_d = v^{10}_d Y_{10^\prime} + v^{126}_d Y_{126}\, , \quad \quad M_u = v^{10}_u Y_{10} + v^{126}_u Y_{126} \, , \nonumber \\
&&M_e = v^{10}_d Y_{10^\prime} - 3 v^{126}_d Y_{126}\, , \quad \quad M^\nu_D = v^{10}_u Y_{10} - 3  v^{126}_u Y_{126} \, , \nonumber \\
&&M_L = v_L Y_{126}\, , \nonumber \\
&&M_R = v_R Y_{126}\, .
\end{eqnarray}
It is convenient to express $Y_{10}$ and $Y_{126}$ in terms of $M_d$ and $M_e$  \cite{Bertolini:2006pe,Buccella:2017jkx} as follows,
\begin{eqnarray}
&&M_u = f_u \left[ \left(3+r \right) M_d + \left( 1-r\right) M_e \right]  \, , \nonumber \\
&&M^\nu_D = f_u \left[3 \left(1-r\right) M_d + \left( 1+ 3 r \right)M_e \right] \, ,
\end{eqnarray}
where
\begin{eqnarray}
f_u = \frac{1}{4} \frac{v^{10}_u}{v^{10}_d}\, , \quad r = \frac{v^{10}_d}{v^{10}_u}  \frac{v^{126}_u}{v^{126}_d}\, . 
\end{eqnarray}
The light neutrino mass formula is governed  by type-II seesaw mechanism, $m^{II}_\nu= M_L$ while type-I seesaw contribution 
$m^{I}_\nu= - M^\nu_D M^{-1}_R M^{\nu^T}_D$ is exactly canceled out 
in the complete diagonalization method. Since light neutrino mass formula is independent of Dirac neutrino mass matrix, any value 
of $M^\nu_D$ is allowed consistent with GUT mass fitting without any fine tuning of the Yukawa couplings. 

\subsection{Discussion on Neutrino Mass}
%
%

The relevant leptonic Yukawa interaction terms for extended seesaw mechanism are given by 
\begin{eqnarray}
-\mathcal{L}_{Yuk} &=& \,\overline{\ell_{L}} \left[Y_3 \Phi + Y_4 \widetilde{\Phi} \right] \ell_R
+ f\, \left[\overline{(\ell_{L})^c} \ell_{L} \Delta_L+\overline{(\ell_{R})^c}\ell_{R}\Delta_R\right] \, \nonumber \\
&&+F\, \overline{(\ell_{R})} H_R S^c_L + F^\prime\, \overline{(\ell_{L})} H_L S_L + \mu_S \overline{S^c_L} S_L\ + \mbox{h.c.}\,. \\
&\supset& M_D \overline{\nu_L} N_R + M_L \overline{\nu^c_L} \nu_L + M_R \overline{N^c_R} N_R \nonumber \\
&&+M \overline{N_R} S_L + \mu_L \overline{\nu^c_L} S_L + \mu_S \overline{S^c_L} S_L
\label{yukaw}
\end{eqnarray}
After spontaneous symmetry breaking, the resulting neutral lepton mass matrix for extended seesaw mechanism in 
the basis $\left(\nu_L, N^c_R, S_L\right)$ is given by
\begin{equation}
\mathbb{M}_\nu= \left( \begin{array}{ccc}
              M_L                   & M_D   & \mu_L  \\
              M^T_D                 & M_R   & M^T \\
              \mu^T_L          & M     & \mu_S
                      \end{array} \right) \, ,
\label{eqn:numatrix}       
\end{equation}
With $\langle H_L \rangle \to 0$ and $\mu_S \to 0$, the complete $9 \times 9$ neutral fermion mass 
matrix in the flavor basis of $\left(\nu_L, S_L, N^c_R \right)$ is read as
\begin{eqnarray}
\mathbb{M} = \left(\begin{array}{c|ccc}   & \nu_L & S_L  & N^c_R   \\ \hline
\nu_L  & M_L       & 0       & M_D \\
S_L    & 0         & 0       & M \\
N^c_R  & M^T_D     & M^T     & M_R
\end{array}
\right).
\label{eq:numatrix-complete}
\end{eqnarray}
Using standard formalism of seesaw mechanism and using mass hierarchy $M_R > M > M_D \gg M_L$, 
we can integrate out the heaviest right-handed neutrinos as follows
\begin{eqnarray}
\mathbb{M}^\prime &=& \begin{pmatrix}
                     M_L & 0 \\
                     0   & 0 
                    \end{pmatrix} 
    -  \begin{pmatrix}
         M_D \\
         M 
        \end{pmatrix} M^{-1}_R
                  \begin{pmatrix}
                   M^T_D & M^T 
                  \end{pmatrix} \nonumber \\
&=&   \begin{pmatrix}
       M_L-M_D M^{-1}_R M^T_D    ~&~ - M_D M^{-1}_R M^T \\
       M M^{-1}_R M^T_D          ~&~  -M M^{-1}_R M^T
       \end{pmatrix} .
       \label{eq:nuS}
\end{eqnarray}
The block diagonalized mass matrices for light left-handed neutrinos, heavy right-handed 
neutrinos and extra sterile neutrinos are
\begin{eqnarray}
&&m_\nu = M_L \,, \nonumber \\
&&M_N \equiv M_R = \frac{v_R}{v_L} M_L\,, \nonumber \\
&&M_S = -M M^{-1}_R M^T \,.
\end{eqnarray}
These block diagonalized mass matrices can be further diagonalized by respective $3 \times 3$ 
unitarity matrices as follows
\begin{eqnarray}
&&m^{\rm diag}_\nu=U^\dagger_\nu m_\nu U^*_\nu = \mbox{diag.}(m_1, m_2, m_3 )\, , \nonumber \\
&&M^{\rm diag}_S=U^\dagger_S M_S U^*_S = \mbox{diag.}( M_{S_1}, M_{S_2}, M_{S_3} )\, , \nonumber \\
&&M^{\rm diag}_N=U^\dagger_N M_N U^*_N = \mbox{diag.}( M_{N_1}, M_{N_2}, M_{N_3} )\,.
\end{eqnarray}
Finally, the complete block diagonalization yields
\begin{eqnarray}
\widehat{\mathbb{M}} &=& \mbox{V}^\dagger_{9 \times 9} \mathbb{M} \mbox{V}^*_{9 \times 9} 
    = \left(\mathbb{W} \cdot \mathbb{U} \right)^\dagger \mathbb{M} \left(\mathbb{W} \cdot \mathbb{U} \right) \nonumber \\
   &=& \mbox{diag.}(m_1, m_2, m_3;\, M_{S_1}, M_{S_2}, M_{S_3}; M_{N_1}, M_{N_2}, M_{N_3} )
\end{eqnarray}
Here the block diagonalized mixing matrix $\mathbb{W}$ and the unitarity matrix $\mathbb{U}$. 

\subsection{Neutrinoless double beta decay from large light-heavy neutrino mixing}
We discuss here the new physics contributions to neutrinoless double beta decay due to large left-right 
neutrino mixing. In the present non-supersymmetric $SO(10)$ GUT  where the TeV scale asymmetric left-right symmetry 
breaking is implemented with both Higgs doublets and triplets, the resulting light neutrino mass is governed by extended 
type-II seesaw mechanism. The type-II seesaw dominance not only provides mass relation between light and heavy neutrinos 
but also allows large Dirac neutrino mass and thereby gives large light-heavy neutrino mixing. This light-heavy neutrino 
mixing plays an important role in giving sizable contributions to neutrinoless double beta decay. 

\begin{figure}[htb!]
\centering
\includegraphics[scale=0.55]{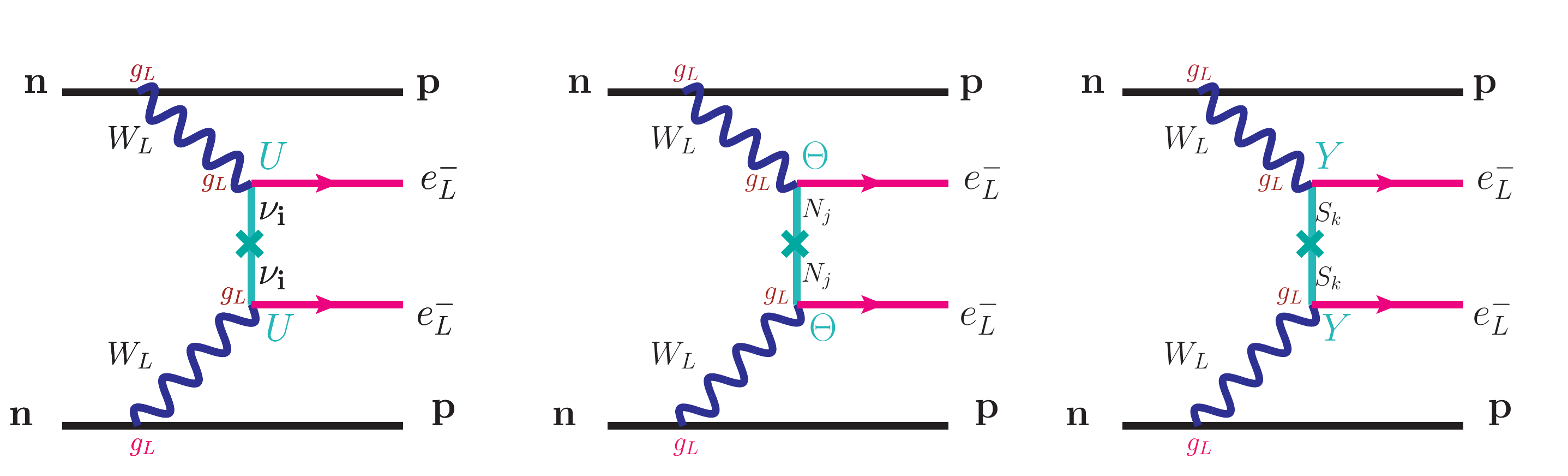}
\caption{Feynman diagrams for neutrinoless double beta decay via $W^-_L - W^-_L$ mediation with the exchange of 
         virtual Majorana neutrinos $\nu_{i}$, $N_j$ and $S_k$.}
\label{feyn:lrsm-WLL}
\end{figure}

The process of neutrinoless double beta decay is governed by following charge current interaction for leptons and quarks 
as following way
\begin{align}
 {\cal L}^{\rm q}_{CC} &=
\frac{g_L}{\sqrt{2}}  \overline{d} \gamma^\mu P_L u W_{L\mu}^- 
+\frac{g_R}{\sqrt{2}} \overline{d}\gamma^\mu P_R u W_{R\mu}^-  + {\rm h.c.} \,, \notag \\
 {\cal L}^{\rm lep}_{CC} &=
 \frac{g_L}{\sqrt{2}} \sum_{\alpha=e, \mu, \tau} \overline{\ell}_{\alpha}\, \gamma^\mu P_L {\nu}_{\alpha }\, W^{-}_{L\mu} 
                      +\frac{g_R}{\sqrt{2}} \sum_{\alpha=e, \mu, \tau} \overline{\ell}_{\alpha}\, \gamma_\mu P_R {N}_{\alpha}\, 
W^{-}_{R\mu} + {\rm h.c.} \,. \notag 
\end{align}
Here the light neutrinos with flavor $\alpha=e, \mu, \tau$ is related with mass eigenstates of neutral leptons as
\begin{align}
&\nu_\alpha=U_{\alpha i} \nu_i + \Theta_{\alpha j} N_j + Y_{\alpha k} S_k    \;.
\end{align}
From the unification plot, it is clear that the mass of right-handed charge gauge boson $W_R$ is kept at very hight scale. 
Because of large value of $M_{W_R}$, the new physics contributions to $0\nu\beta\beta$ due to purely right-handed current 
is negligible as these contributions are proportional to $1/M^4_{W_R}$. Similarly the well known $\lambda-$ ($\propto 1/M^2_{W_R}$) 
and $\eta-$  ($\propto \tan \theta^W_{LR}$) diagrams are very much suppressed. The only contributions which can contribute 
to $0\nu\beta\beta$ are arising from purely left-handed currents due to exchange of light left-handed neutrinos $\nu_L$,  right-handed 
neutrinos $N_R$ and sterile neutrinos $S_L$ as displayed in Fig.    \ref{feyn:lrsm-WLL}.

%

Since the flavor neutrino state $\nu_\alpha$ is a linear combination of mass eigenstates 
$\nu_i, N_j, S_k$, there are three different Feynman diagrams contributing to neutrinoless 
double beta decay arising from purely left-handed charge current interaction as shown in Fig. \ref{feyn:lrsm-WLL}. 
The Feynman amplitude for these diagrams are proportional to
\begin{eqnarray}
\label{eq:amp_LL} 
& &\mathcal{A}_{LL}^{\nu} \propto G^2_F \sum_{i=1,2,3} \frac{U_{ei}^2\, 
                      m_{\nu_i}}{p^2} \,, \nonumber \\
& &\mathcal{A}_{LL}^{N} \propto G^2_F \sum_{j=1,2,3} \left(-\frac{\Theta_{ej}^2}{M_{N_j}} \right)\,,\nonumber \\
& &\mathcal{A}_{LL}^{S} \propto G^2_F \sum_{k=1,2,3} \left(-\frac{Y_{ek}^2}{M_{S_k}} \right)\,,                       
\end{eqnarray}
The resulting effective Majorana mass parameters, a measure of lepton number violation for these contributions 
are given by
\begin{eqnarray}
\label{eq:eta_LL} 
& &|\mee^{\nu}| = \sum_{i=1,2,3} U_{ei}^2\, m_{\nu_i} \,, ~~
  |\mee^{N}| = \langle p^2 \rangle \sum_{j=1,2,3} \frac{\Theta_{ej}^2}{M_{N_j}} \,,  ~~                 
  |\mee^{S}| = \langle p^2 \rangle \sum_{k=1,2,3} \frac{Y_{ek}^2}{M_{S_k}} \,. \nonumber
\end{eqnarray}
The corresponding inverse half-life expression for this rare process $0\nu\beta\beta$ due to exchange of 
$\nu$, $N$ and $S$ is given as follows
\begin{eqnarray}
\left[T_{1/2}^{0\nu}\right]^{-1}&=&\mathcal{K}_{0\nu}\, 
    \bigg\{ |{\large \bf  m}_{\rm ee}^{\nu}|^2 + |{\large \bf m}_{\rm ee}^{S}+{\large \bf m}_{\rm ee}^{N}|^2 \bigg\}\;,
\end{eqnarray}
where $\mathcal{K}_{0\nu}\simeq 1.57 \times 10^{-25} {\rm yrs}^{-1} {\rm eV}^{-2}$ is the product of the phase space factor, square of the nuclear matrix elements (NME) and the inverse of electron mass square.
\begin{table}[t]
 \centering
\vspace{10pt}
 \begin{tabular}{c|c}
 \hline \hline
Effective Mass Parameters & Analytic formula  \\
\hline
${\large \bf  m}_{\rm ee,L}^{\nu}$ & $\sum_{i=1}^3 U_{e\,i}^2\, m_{\nu_i} $  \\
${\large \bf  m}_{\rm ee,L}^{N}$   & $\sum_{i=1}^3 \Theta_{e\,i}^2\, \frac{|p|^2}{M_{N_i}}$  \\
${\large \bf  m}_{\rm ee,L}^{S}$   & $\sum_{i=1}^3 Y_{e\,i}^2\, \frac{|p|^2}{M_{S_i}}$   \\
\hline \hline
 \end{tabular}
 \caption{The effective Majorana mass parameters arising from purely left-handed currents due to 
 exchange of left-handed neutrinos $\nu_L$, right-handed neutrinos $N_R$ and sterile neutrinos $S_L$.}
 \label{tab:mee_LL}
\end{table}

\subsection{Numerical result for $0\nu\beta\beta$}
We aim to present numerically different new physics contributions to $0\nu\beta\beta$ transition with the variation 
of lightest neutrino mass and derive lower bound on lightest neutrino mass by saturating the existing experimental 
bound on $0\nu\beta\beta$. The input model parameters needed for the said purpose are listed as follows:
\begin{itemize}
\item  {\bf Structure of $M_D$}\\
Since $SO(10)$ GUT contains Pati-Salam symmetry as highest subgroup relating quarks and leptons, the structure of 
Dirac neutrino mass matrix is expected to be of the order of up-type quark mass matrix. Assuming TeV scale asymmetric 
left-right symmetry originated from Pati-Salam symmetry~\cite{Pati:1974yy} or $SO(10)$ GUT~\cite{Fritzsch:1974nn}, 
and without considering RG corrections, the form of $M_D$ is derived to be\footnote{One can go through refs.~\cite{Parida:2014dba,Dev:2009aw,LalAwasthi:2011aa,Awasthi:2013ff}  
where RG effect has been taken into account in constructing the Dirac neutrino mass matrix at TeV scale.}
\begin{eqnarray}
M_D&=& V_{CKM}\,M_u\,V^{T}_{CKM} \nonumber \\
&=&\left(
\begin{array}{ccc}
 0.067-0.004\,i ~&~ 0.302-0.022\,i ~ & ~0.550-0.530\,i \\
 0.302-0.022\,i & 1.480           & 6.534-0.001\,i \\
 0.550-0.530\,i ~&~ 6.534-0.0009\,i~ &~ 159.72
\end{array}
\right)\text{GeV}\, , \nonumber
\end{eqnarray}
where the PDG~~\cite{Agashe:2014kda} value of up-type quark mass matrix and the CKM mixing matrix are given as 
follows
\begin{eqnarray}
&&M_u= \mbox{diag}(2.3~\mbox{MeV},1.275~\mbox{GeV},173.210~\mbox{GeV})\, , \nonumber \\
&&V_{CKM}=
\begin{pmatrix}
0.97427 & 0.22534 & 0.00351-i0.0033 \\
-0.2252+i0.0001 & 0.97344 & 0.0412 \\
0.00876-i0.0032 ~&~ -0.0404-i0.0007 ~&~ 0.99912 
\end{pmatrix}\, .
\end{eqnarray}
\item  {\bf Structure of $M$}\\
The $N_R-S_L$ mixing matrix $M$ is assumed to be diagonal and degenerate for simplicity of the numerical calculations and the mass matrix 
is given by
\begin{eqnarray}
M=m_0
\begin{pmatrix}
1 & 0 & 0 \\
0 & 1 & 0  \\
0 & 0 & 1 
\end{pmatrix}\, ,
\end{eqnarray}
where we chose $m_0$ to be $200~$GeV,  $500~$GeV and $1000~$GeV for our numerical estimations.

\item {\bf Physical masses and mixing for neutral leptons}\\
The complete diagonalization of neutral lepton in LRSM with additional fermion singlet has been discussed 
in Ref.~\cite{Pritimita:2016fgr} and we summarise them for our present discussion. The physical masses 
for light left-handed Majorana neutrinos, heavy right-handed Majorana neutrinos and sterile neutrinos in 
terms of oscillation parameters is given by
\begin{eqnarray}
& &m_\nu = U_{\rm PMNS} m^{\rm diag}_\nu U^T_{\rm PMNS} \, , \nonumber \\[2mm]
& &M_N \equiv M_R = \frac{v_R}{v_L} U_{\rm PMNS} m^{\rm diag}_\nu U^T_{\rm PMNS} \,, \nonumber  \\[1mm]
& &M_S = -m^2_0\, \frac{v_L}{v_R} U^*_{\rm PMNS} {m^{\rm diag}_\nu}^{-1} U^\dagger_{\rm PMNS} \,.
\end{eqnarray}

The relevant mixing elements needed for $0\nu\beta\beta$ are like light-light active neutrino mixing matrix as 
$U_\nu = U_{\rm PMNS}$,  mixing between left-handed and right-handed neutrinos as $\Theta = 
\frac{v_L}{v_R} M_D U^{-1}_{\rm PMNS} {m^{\rm diag}_{\nu}}^{-1}$ and mixing element between light active 
neutrinos and sterile neutrinos as $Y=\frac{1}{m_0} M_D U^*_{\rm PMNS}$.
\item {\bf $U_{\rm PMNS}$ and Oscillation Parameters}

The light active neutrino matrix can be  diagonalized by the Pontecorvo-Maki-Nakagawa-Sakata (PMNS) mixing matrix $U_{\rm PMNS}$ as 
$$m^{\rm diag}_\nu=U^\dagger_{\rm PMNS} m_\nu U^*_{\rm PMNS}= \mbox{diag.}(m_1, m_2, m_3)$$ 
where
\begin{eqnarray}
U_{\rm {PMNS}}&=& \begin{pmatrix} c_{13}c_{12}&c_{13}s_{12}&s_{13}e^{-i\delta}\\
-c_{23}s_{12}-c_{12}s_{13}s_{23}e^{i\delta}&c_{12}c_{23}-s_{12}s_{13}s_{23}e^{i\delta}&s_{23}c_{13}\\
s_{12}s_{23}-c_{12}c_{23}s_{13}e^{i\delta}&-c_{12}s_{23}-s_{12}s_{13}c_{23}e^{i\delta}&c_{13}c_{23}
\end{pmatrix}\cdot \mbox{P}\,.
\label{PMNS} 
\end{eqnarray}
We denote $s_{ij}=\sin \theta_{ij}$, $c_{ij}=\cos \theta_{ij}$ and diagonal phase matrix $\mbox{P}=\mbox{diag}\left(1, e^{i\alpha}, e^{i \beta} \right)$, 
where $\delta$ is the Dirac CP phase and $\alpha$, $\beta$ are Majorana phases. We vary them from $0$ to $2 \pi$ in our numerical computation. 
The mixing angles like the atmospheric mixing angle as $\theta_a\equiv \theta_{23}$, the solar mixing angle as $\theta_s\equiv \theta_{12}$, 
and the reactor mixing angle as $\theta_r\equiv \theta_{13}$ and their $3\sigma$ ranges are presented in Table.\,\ref{table-osc} along with the 
two mass squared differences like ($\Delta\,m_{\rm atm}^2$) and ($\Delta\,m_{\rm sol}^2$). 
Since oscillation experiments unable to provide precise measurement of the sign of $\Delta\,m_{\rm atm}^2$, there 
can be two possibilities in the arrangement of light neutrino masses like
      
\textbf{Normal hierarchy (NH)}: $\Delta\,m_{\rm atm}^2 \equiv \Delta m_{31}^2 > 0$, which gives $ m_1 < m_2 < m_3$ with
      \begin{equation*}
         m_2 = \sqrt{m_1^2 +\Delta\,m_{\rm sol}^2}\;,\qquad 
         m_3 = \sqrt{m_1^2 +\Delta\,m_{\rm atm}^2}\;, \label{eq1:NH_m2_m3}
      \end{equation*}
  
\textbf{Inverted hierarchy (IH)}: $\Delta\,m_{\rm atm}^2 \equiv \Delta m_{31}^2 < 0$, implying $m_3 < m_1 < m_2$ with
      \begin{equation*}
         m_1 = \sqrt{m_3^2 +\Delta\,m_{\rm atm}^2}\;,\qquad 
         m_2 = \sqrt{m_3^2 +\Delta\,m_{\rm atm}^2+\Delta\,m_{\rm sol}^2}\;. \label{eq1:IH_m1_m2}
      \end{equation*}
      
\begin{table}[htb!]
\begin{tabular}{cccc}
        \hline 
Oscillation Parameters &  3$\sigma$ range         &  3$\sigma$ range &  3$\sigma$ range  \\
           &   (Schwetz et al.\cite{GonzalezGarcia:2012sz})   &   (Fogli et al.\cite{Fogli:2012ua}) & (Gonzalez-Garcia et al 
           \cite{Gonzalez-Garcia:2014bfa}) \\
        \hline \hline
$\Delta m^2_{\rm {21}} [10^{-5}\mbox{eV}^2]$              & 7.00-8.09   & 6.99-8.18  & 7.02 - 8.09   \\
$|\Delta m^2_{\rm {31}}(\mbox{NH})| [10^{-3} \mbox{eV}^2]$ & 2.27-2.69   & 2.19-2.62  & 2.317 - 2.607   \\
$|\Delta m^2_{\rm {31}}(\mbox{IH})| [10^{-3} \mbox{eV}^2]$ & 2.24-2.65   & 2.17-2.61  & 2.307 - 2.590   \\
\hline
$\sin^2\theta_{s}$                                        & 0.27-0.34   & 0.259-0.359 & 0.270 - 0.344  \\
$\sin^2\theta_{a}$                                        & 0.34-0.67   & 0.331-0.637 & 0.382 - 0.643  \\
$\sin^2\theta_{r}$                                        & 0.016-0.030 & 0.017-0.031 & 0.0186 - 0.0250  \\
        \hline
\end{tabular}
\caption{The mass squared differences and mixing angles with their allowed $3\sigma$ range. However we have taken the values 
              given in Ref.\cite{Gonzalez-Garcia:2014bfa}  for our numerical computation.}
\label{table-osc}
\end{table}

In this work where the light neutrino mass is governed by type-II seesaw dominance, the heavy neutrino mass eigenvalues 
are related to light neutrino mass eigenvalues as follows:

For NH pattern
   \begin{eqnarray}
    && M_{N_1}=M_1 = \frac{m_{1}}{m_{3}}\, M_{3}\,,\quad 
       M_{N_2}=M_2 = \frac{m_{2}}{m_{3}}\, M_{3}\,.
   \end{eqnarray}
 where $M_{3}$ being the largest mass eigenvalue for right-handed Majorana neutrinos. 
 
 For IH pattern 
   \begin{eqnarray}
   && M_{N_1}=M_1 = \frac{m_{1}}{m_{2}}\, M_{2}\,,\quad 
      M_{N_3}=M_3 = \frac{m_{3}}{m_{2}}\, M_{2}\,. 
   \end{eqnarray}
$M_{2}$ being largest mass eigenvalue of right-handed neutrinos. We have fixed $M_3 (M_2)$ as $5~$TeV for NH (IH) 
pattern.  
\end{itemize}

The expression for inverse half-life for a given isotope for these contributions due to exchange of light neutrinos, heavy right-handed neutrinos and 
heavy sterile neutrinos is given by
\begin{align}
 [T_{1/2}^{0\nu}]^{-1} \!=\! G_{01} 
 \left(|\mathcal{M}_\nu \eta_{\nu}|^2 + |\mathcal{M}_N \left(\eta_N +  \eta_S\right)|^2\right),
\end{align}
where $G_{01}$ represents standard $0\nu\beta\beta$ phase space factor, $\mathcal{M}_i$ as the corresponding nuclear matrix elements for the different exchange processes 
and $\eta_i$ are dimensionless particle physics parameters. 

\subparagraph*{Due to Light neutrinos:} 
The lepton number violating dimensionless particle physics parameter for 
$0\nu\beta\beta$ transition due to the exchange of light neutrinos is given by
\begin{align}
\label{eta:nu} 
\mathcal{\eta}_{\nu} =\frac{1}{m_e}  \sum^{3}_{i=1} U^2_{ei}\, m_{i}
           = \frac{m^\nu_{\rm ee}}{m_e} \,,
\end{align}
where $m_e$ is the electron mass. 

\begin{figure*}[htb!]
\includegraphics[width=0.48\textwidth]{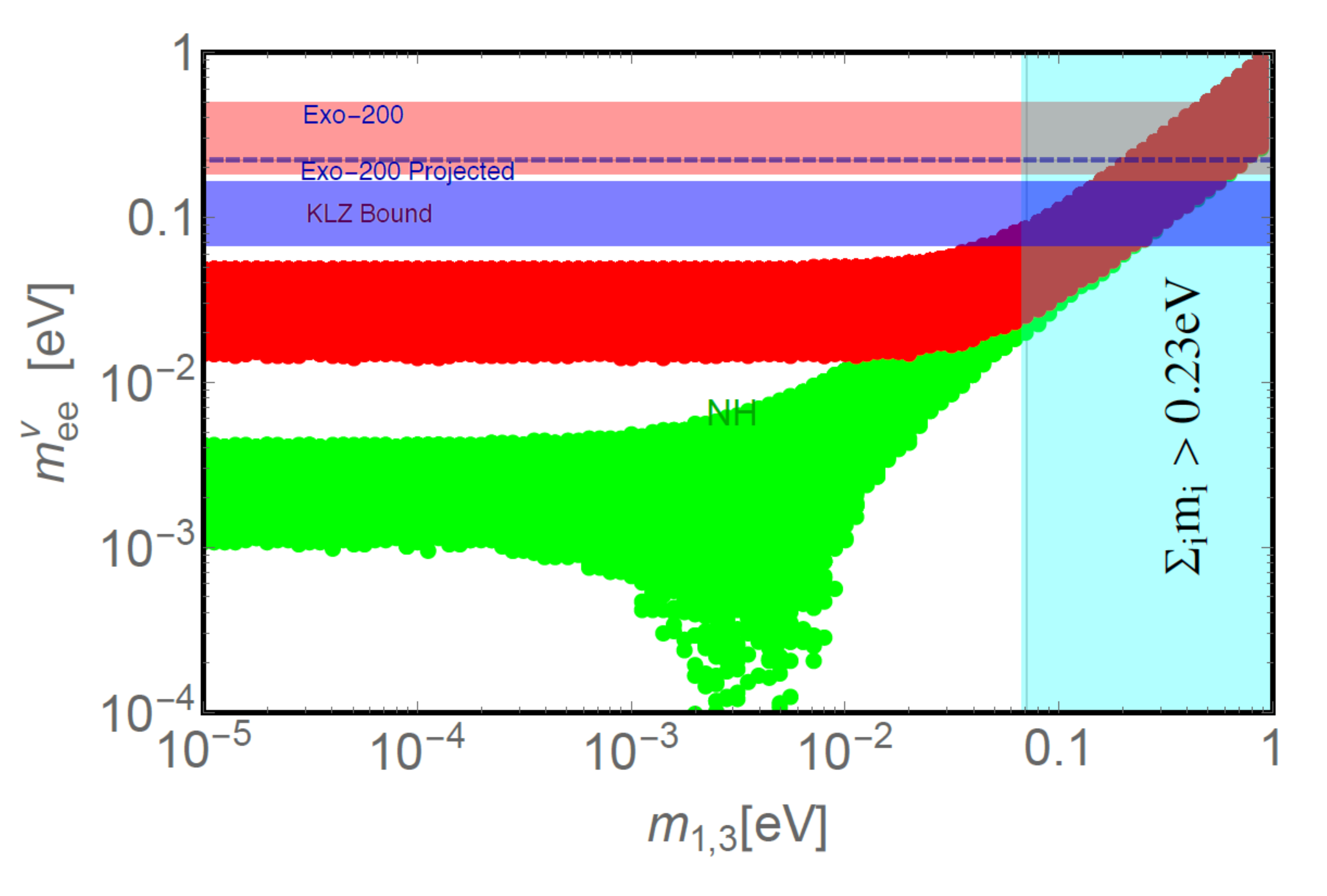}
\hspace*{0.2 truecm}
\includegraphics[width=0.48\textwidth]{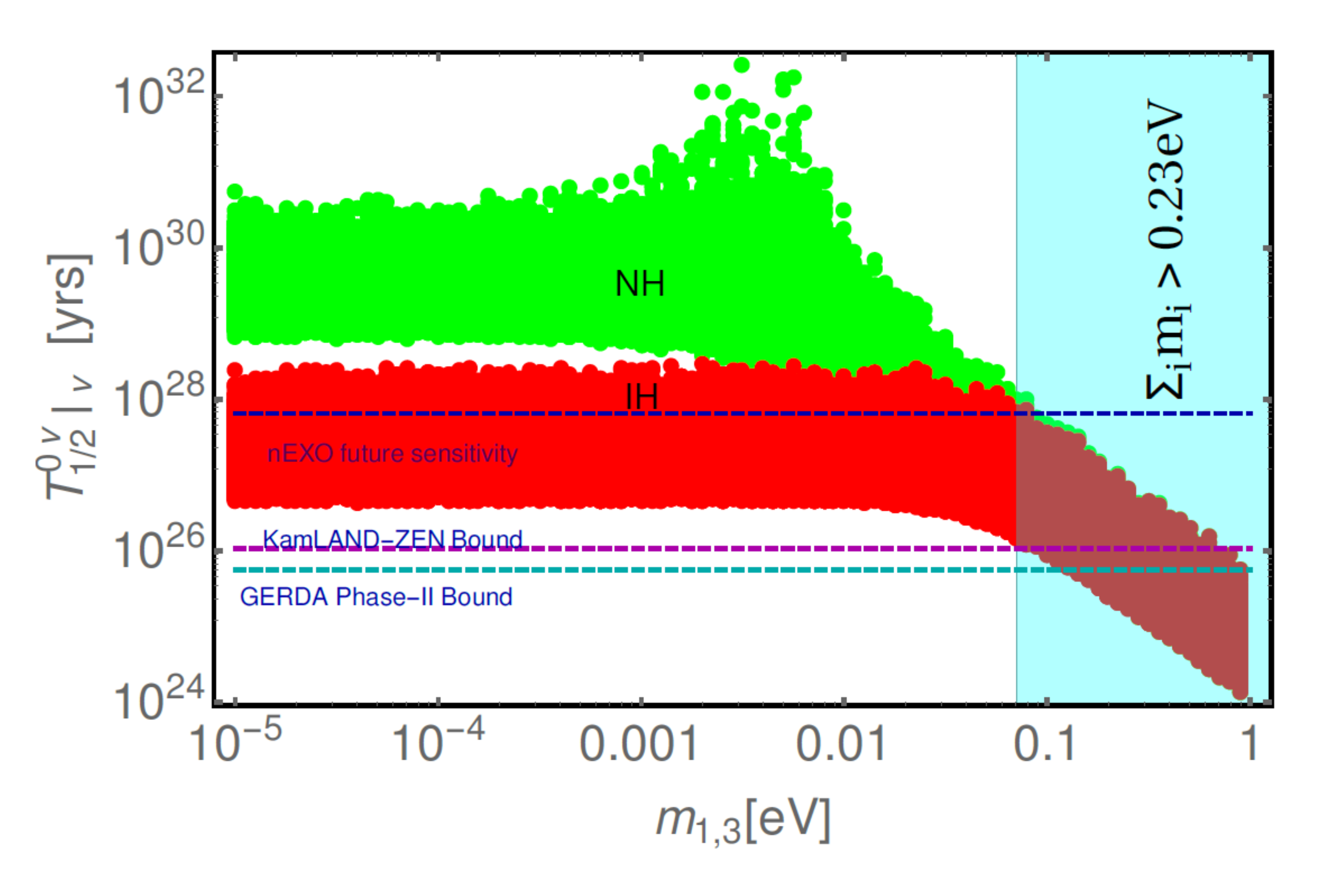}
\caption{Effective Majorana mass parameter and Half-life of $0\nu\beta\beta$ decay 
         due to exchange of light active neutrinos as a function of the lightest neutrino mass for a NH and IH pattern.} 
\label{plot:0nubb-1}
\end{figure*}
The corresponding effective mass parameter is 
given by
\begin{align}
\label{eq:mee-std}
m^\nu_{\rm ee}
=\left| c^2_{12} c^2_{13} m_1 + s^2_{12} c^2_{13} m_2 e^{i\alpha} + s^2_{13} m_3 e^{i\beta} \right| \,,
\end{align}
with $\theta_{12}$ and $\theta_{13}$, $c_{12} = \cos\theta_{12}$, etc as the respective sine and cosine of the oscillation angles. The two unconstrained Majorana phases varied between $0 \leq \alpha,\beta < 2\pi$.

\subparagraph*{Due to right-handed neutrinos:}
The new non-standard contribution to $0\nu\beta\beta$ decay arising from the purely left-handed 
currents via the exchange of right-handed neutrinos results in the lepton number violating dimensionless particle physics parameter
\begin{align}
\label{eta:N} 
\mathcal{\eta}_N &= 
	m_p \sum^{3}_{i=1} \frac{\Theta^2_{ei} M_i}{|p|^2 + M^2_i}\,.
\end{align}
We define here $p$ as the virtual neutrino momentum of the order of the nuclear 
Fermi scale, $p\approx 100$~MeV and $m_p$ is the proton mass. 

\begin{figure*}[t!]
\includegraphics[width=0.48\textwidth]{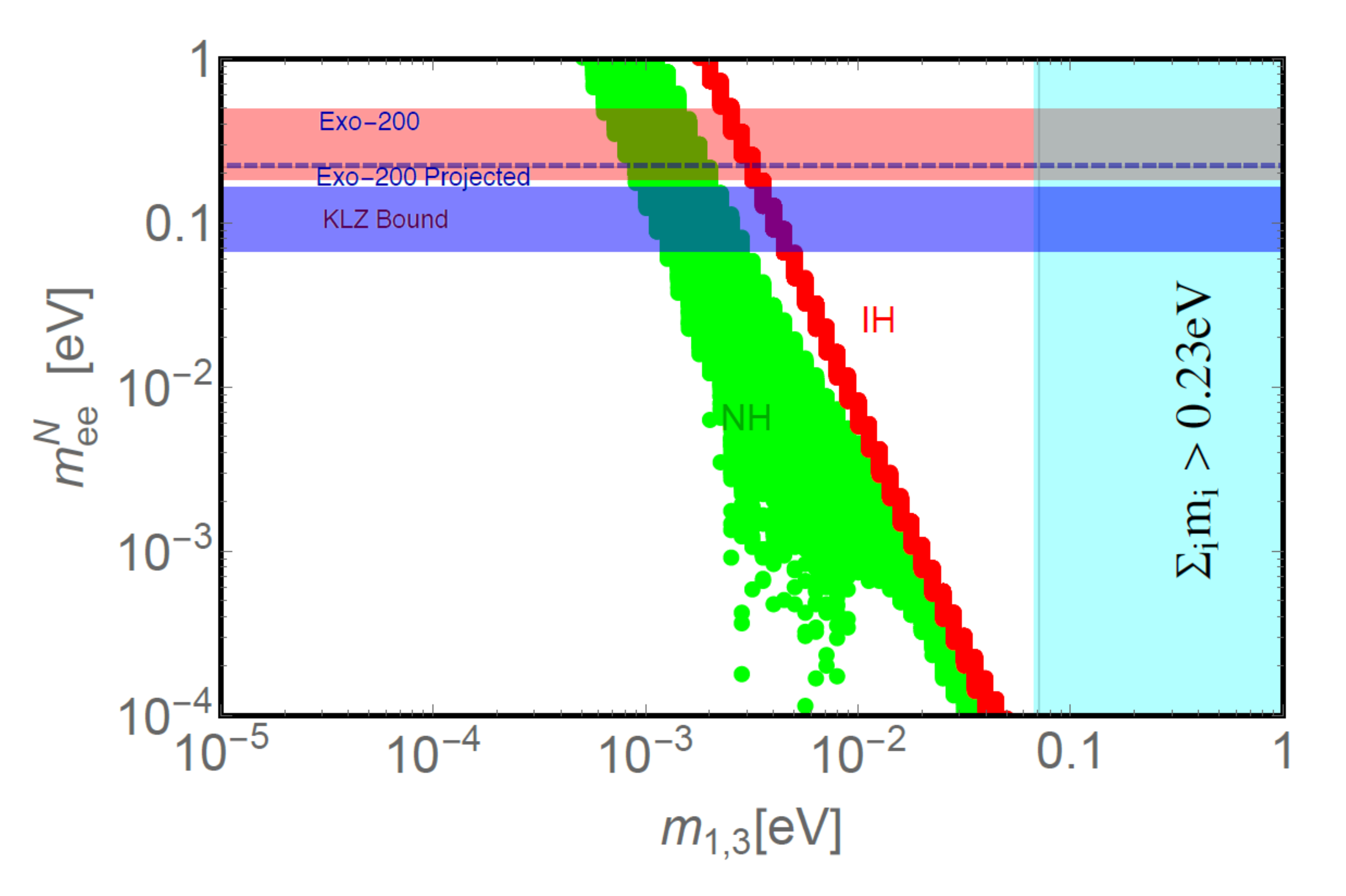}
\hspace*{0.2 truecm}
\includegraphics[width=0.48\textwidth]{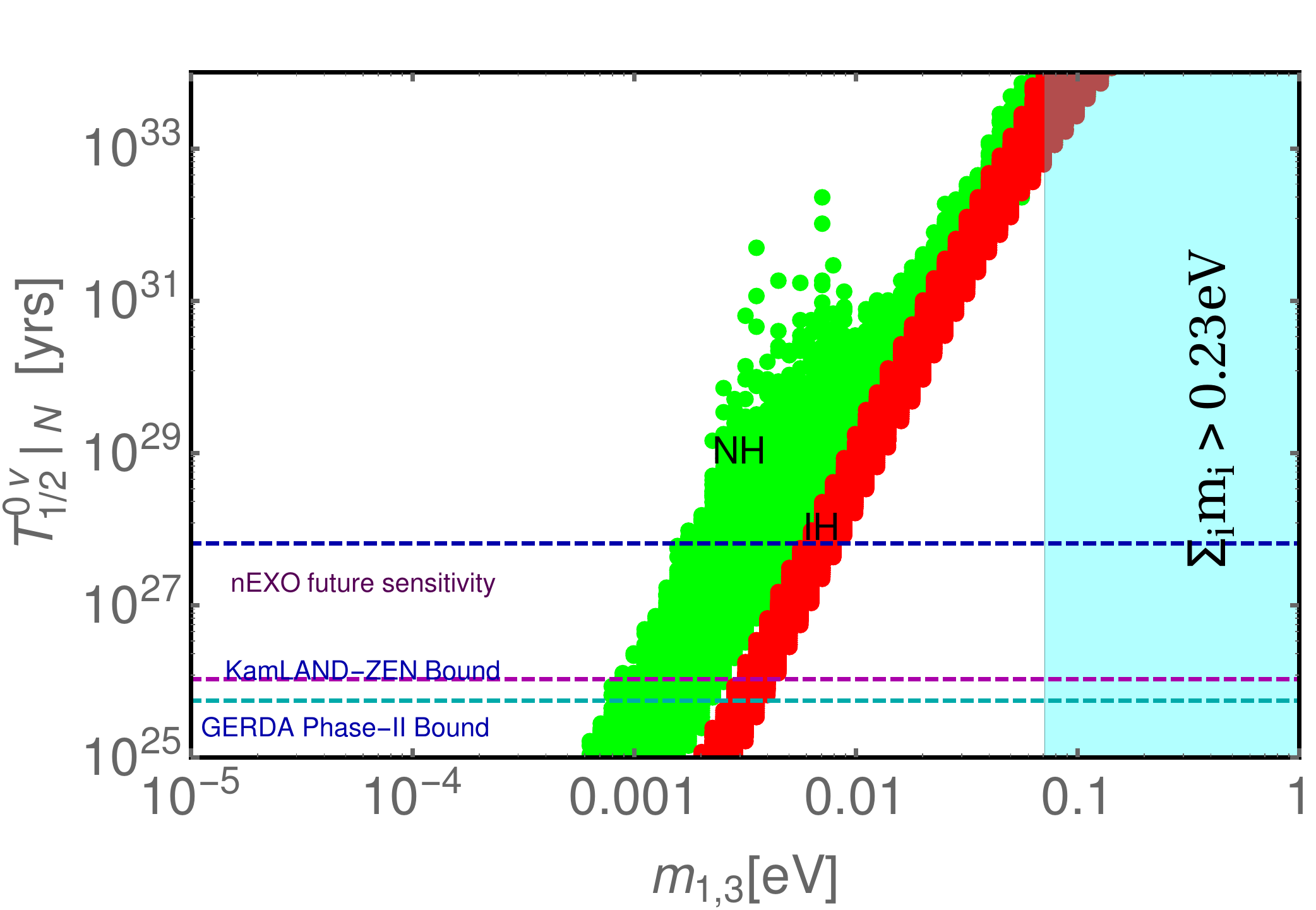}
\caption{Effective Majorana mass parameter and Half-life of $0\nu\beta\beta$ decay 
         due to exchange of heavy right-handed neutrinos as a function of the lightest neutrino mass for a NH and IH pattern.} 
\label{plot:0nubb-2}
\end{figure*}
We also consider 
mass of right-handed neutrinos larger than nuclear fermi scale i.e, $M^2_i \gg |p^2|$. 
With $M_i \gg |p|$, the propagator simplifies as 
\begin{align}
\frac{M_i}{p^2-M_i^2} \approx -\frac{1}{M_i}\,.
\end{align}
Thus, the effective dimensionless particle physics parameter due to exchange of heavy right-handed neutrinos 
is found to be
\begin{align}
\label{eta:Na} 
\mathcal{\eta}_N &= 
	- m_p \sum^{3}_{i=1} \frac{\Theta^2_{ei} }{M_{i}}\,, 
\end{align}
and the corresponding effective Majorana mass parameter is given by
\begin{align}
\label{mee:N} 
m^N_{\rm ee} &= 
	 \sum^{3}_{i=1} \frac{\Theta^2_{ei} }{M_{i}}  \langle p^2 \rangle \,.
\end{align}
\begin{figure*}[htb!]
\includegraphics[width=0.48\textwidth]{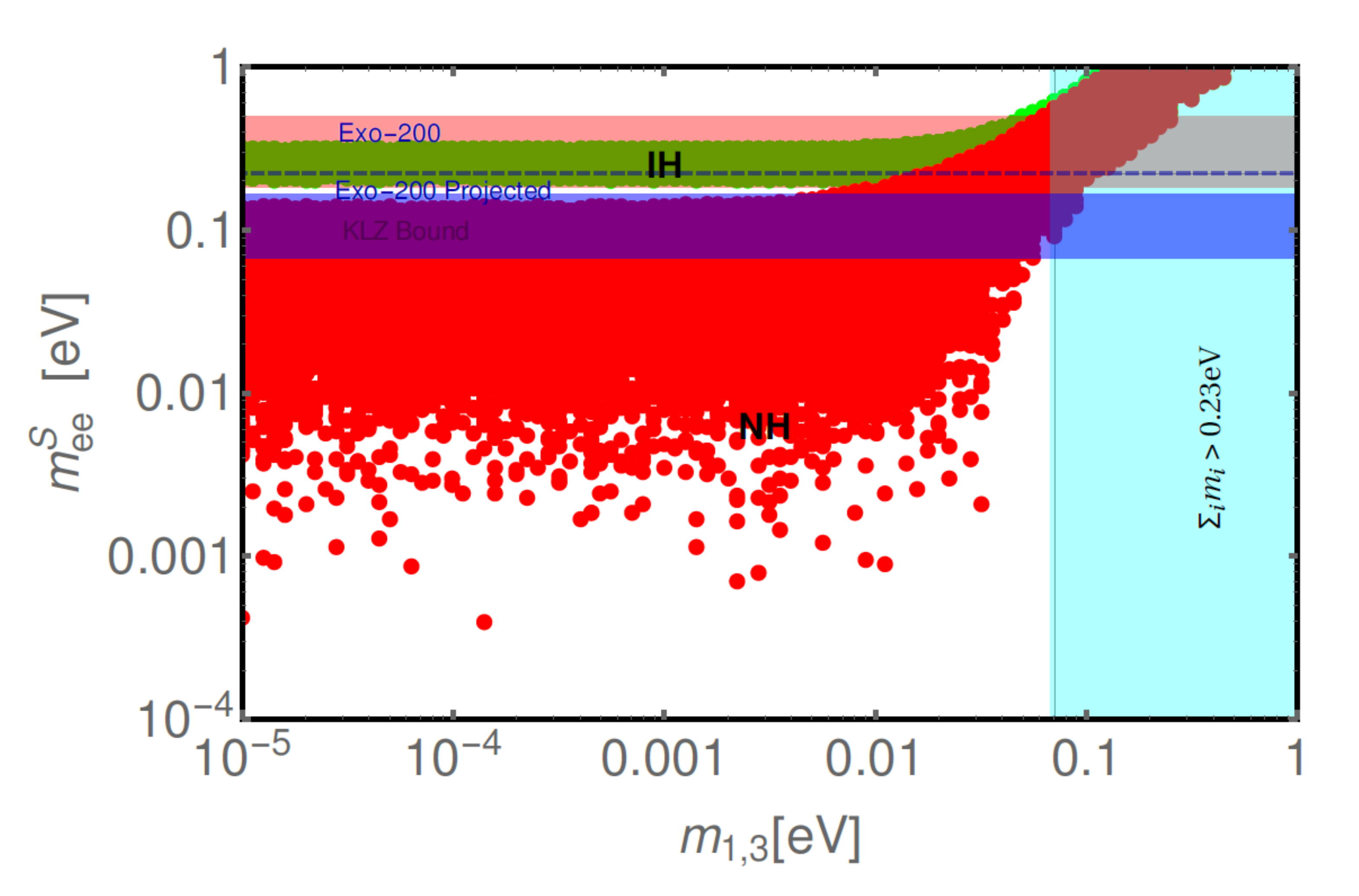}
\hspace*{0.2 truecm}
\includegraphics[width=0.48\textwidth]{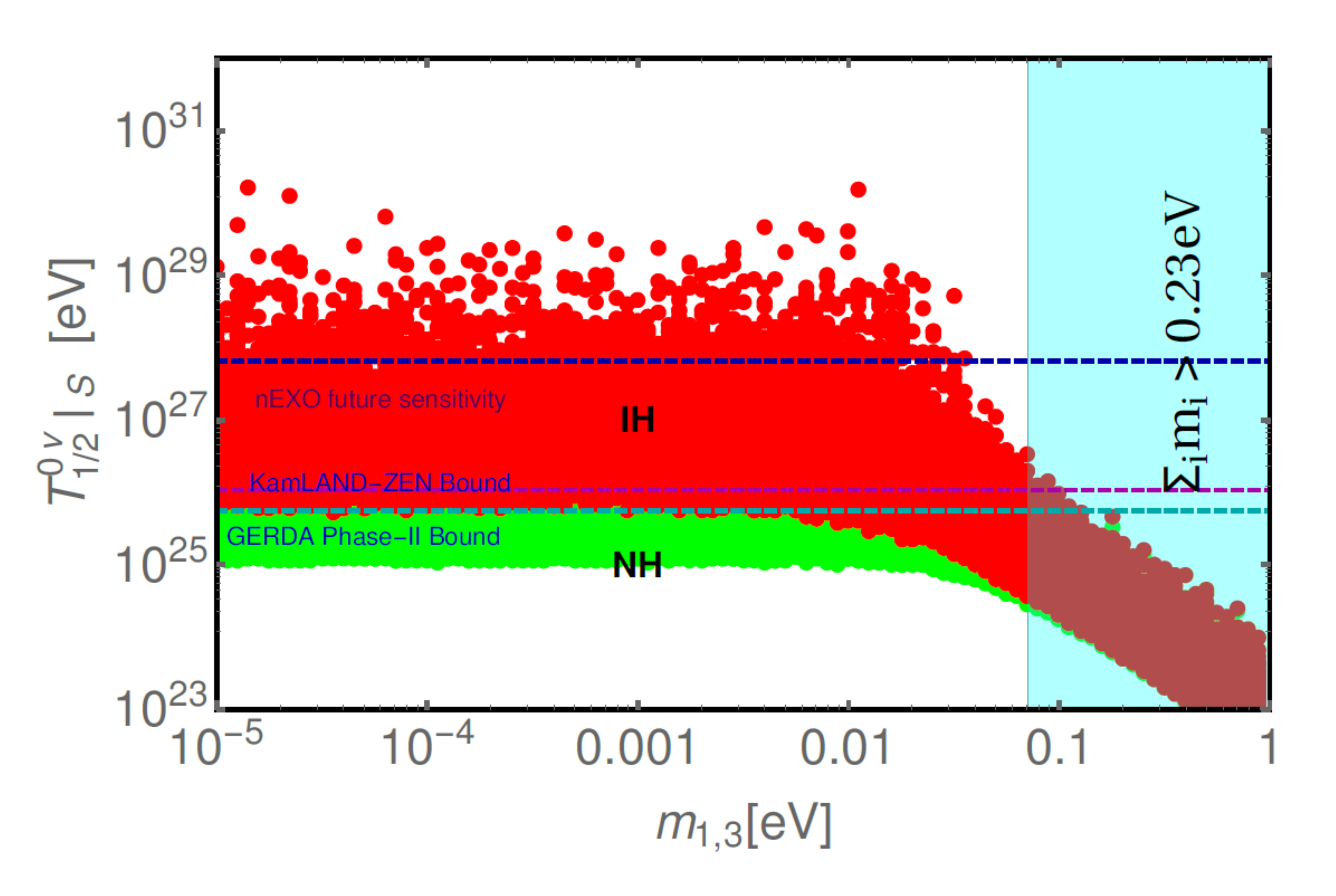}
\caption{Effective Majorana mass parameter and Half-life of $0\nu\beta\beta$ decay 
         due to exchange of heavy sterile neutrinos as a function of the lightest 
         neutrino mass for a NH and IH pattern.} 
\label{plot:0nubb-3}
\end{figure*}
\subparagraph*{Due to heavy sterile neutrinos:}
Similarly, the contribution to $0\nu\beta\beta$ decay arising from heavy sterile neutrinos gives lepton number violating dimensionless particle physics parameter 
as,
\begin{align}
\label{eta:N} 
\mathcal{\eta}_S &= 
	m_p \sum^{3}_{i=1} \frac{Y^2_{ei} M_{S_i}}{|p|^2 + M^2_{S_i}}\,.
\end{align}
Assuming $M_{S_i} \gg |p|$, the dimensionless particle physics parameter and effective Majorana mass parameter 
due to exchange of sterile neutrinos are expressed as,
\begin{align}
\label{mee:S} 
\mathcal{\eta}_S &=  - m_p  \sum^{3}_{i=1} \frac{Y^2_{ei} }{M_{S_i}}  \, , \quad \quad 
m^S_{\rm ee} = 
	 \sum^{3}_{i=1} \frac{Y^2_{ei} }{M_{S_i}}  \langle p^2 \rangle \,.
\end{align}

\begin{figure*}[htb!]
\includegraphics[width=0.48\textwidth]{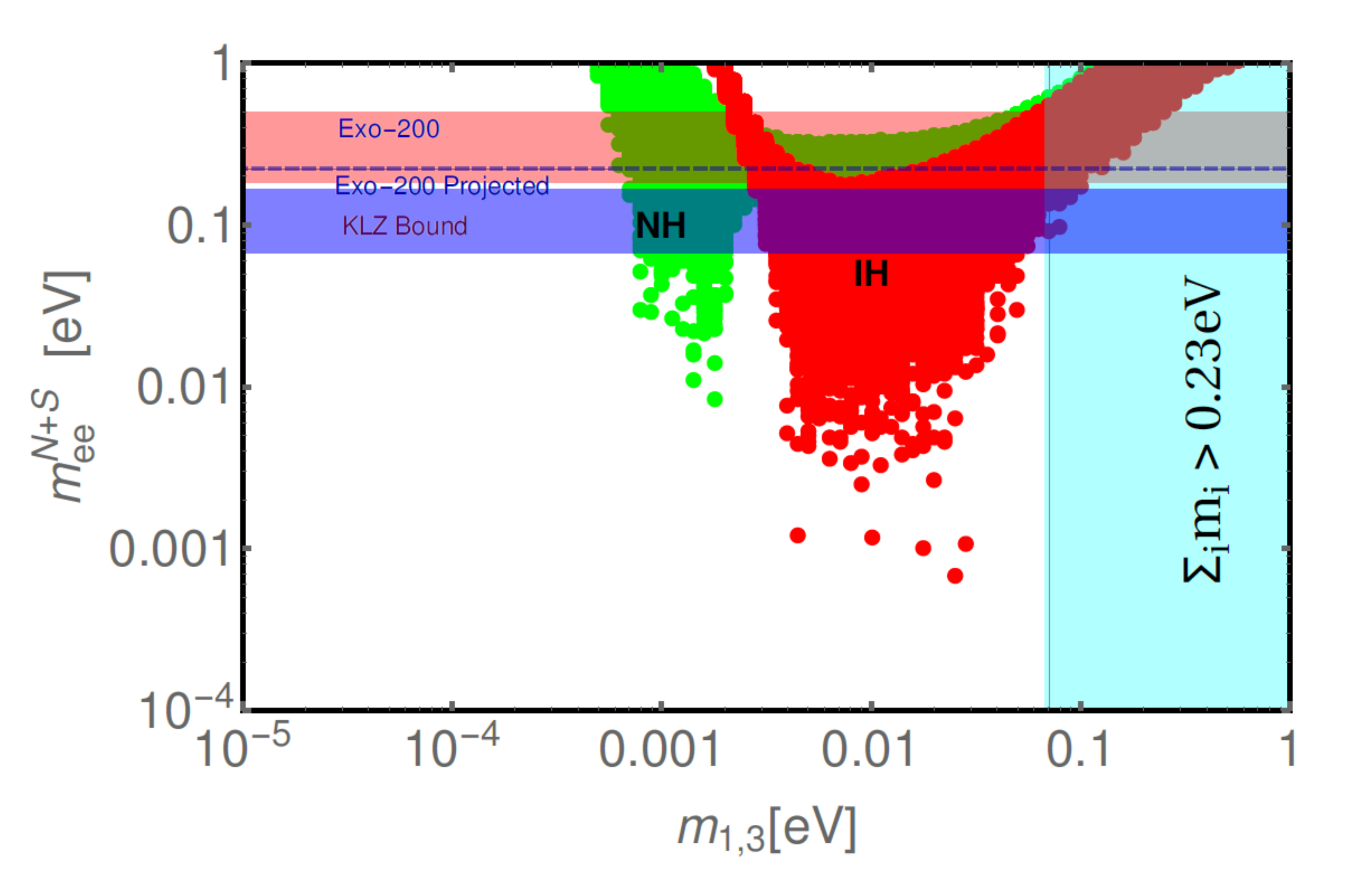}
\hspace*{0.2 truecm}
\includegraphics[width=0.48\textwidth]{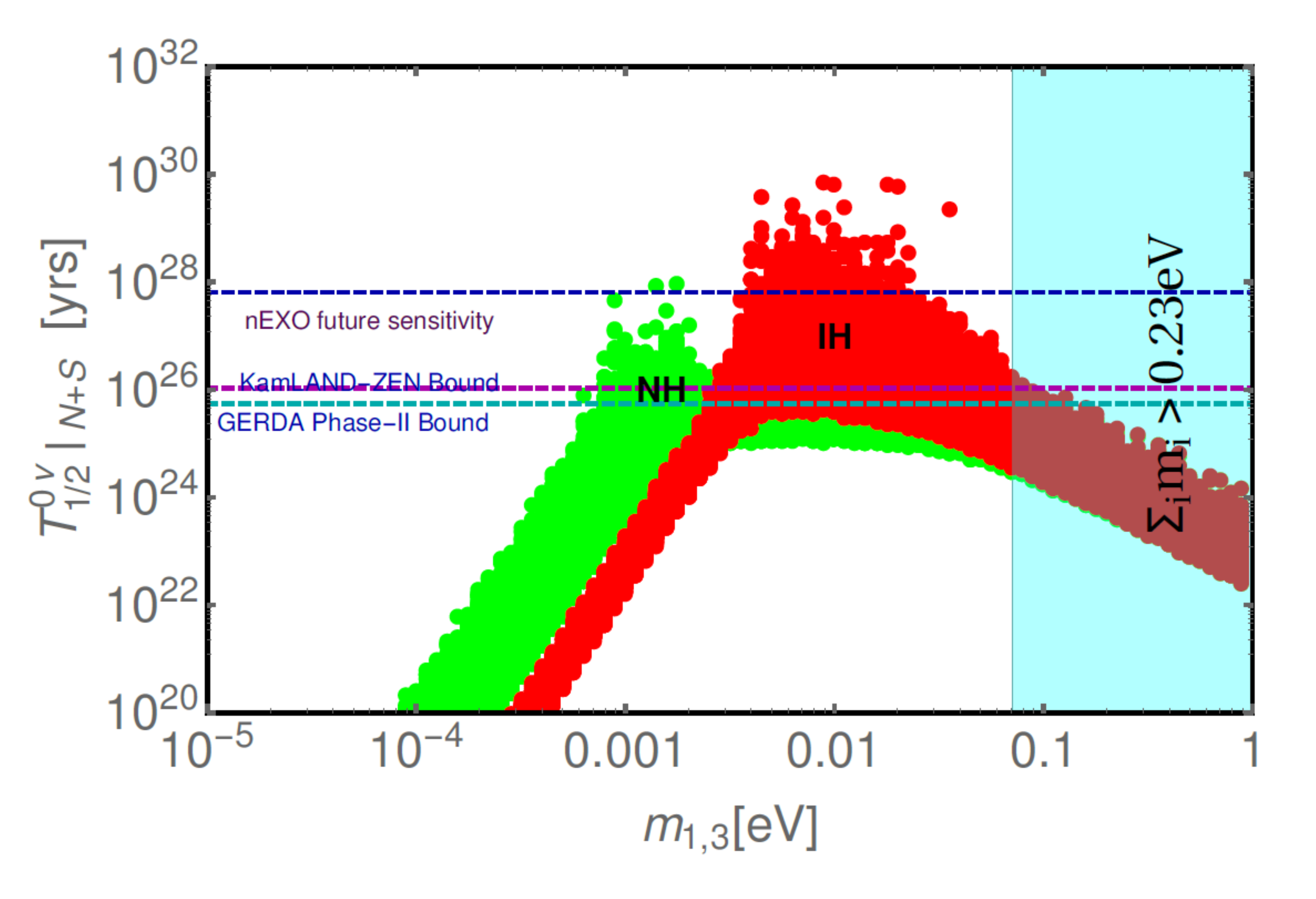}
\caption{Effective Majorana mass parameter and Half-life of $0\nu\beta\beta$ decay 
         due to combine effect of heavy right-handed neutrinos plus sterile neutrinos 
         as a function of the lightest neutrino mass for a NH and IH pattern.} 
\label{plot:0nubb-4}
\end{figure*}
\subparagraph*{Combined contributions}
The combined contribution to effective Majorana mass parameter due to exchange of 
heavy right-handed neutrinos as well as sterile neutrinos is given by
\begin{align}
\label{mee:S} 
m^{N+S}_{\rm ee} =   \sum^{3}_{i=1} \frac{\Theta^2_{ei} }{M_{i}}  \langle p^2 \rangle +
	 \sum^{3}_{i=1} \frac{Y^2_{ei} }{M_{S_i}}  \langle p^2 \rangle \,.
\end{align}

\begin{figure}[!htb]
\begin{center}
\includegraphics[width=7cm,height=5cm]{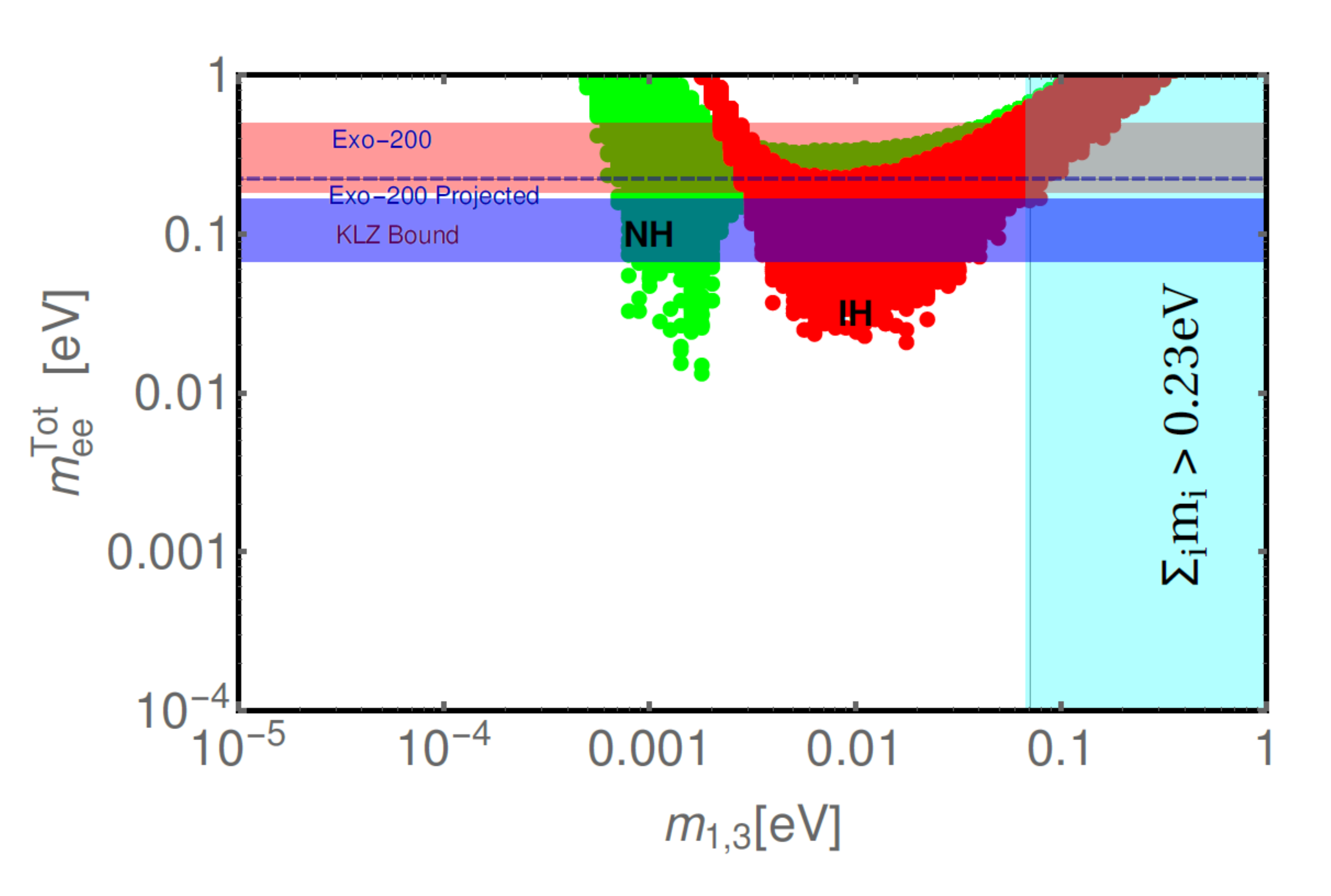}
\hspace*{0.2 truecm}
\includegraphics[width=7cm,height=5cm]{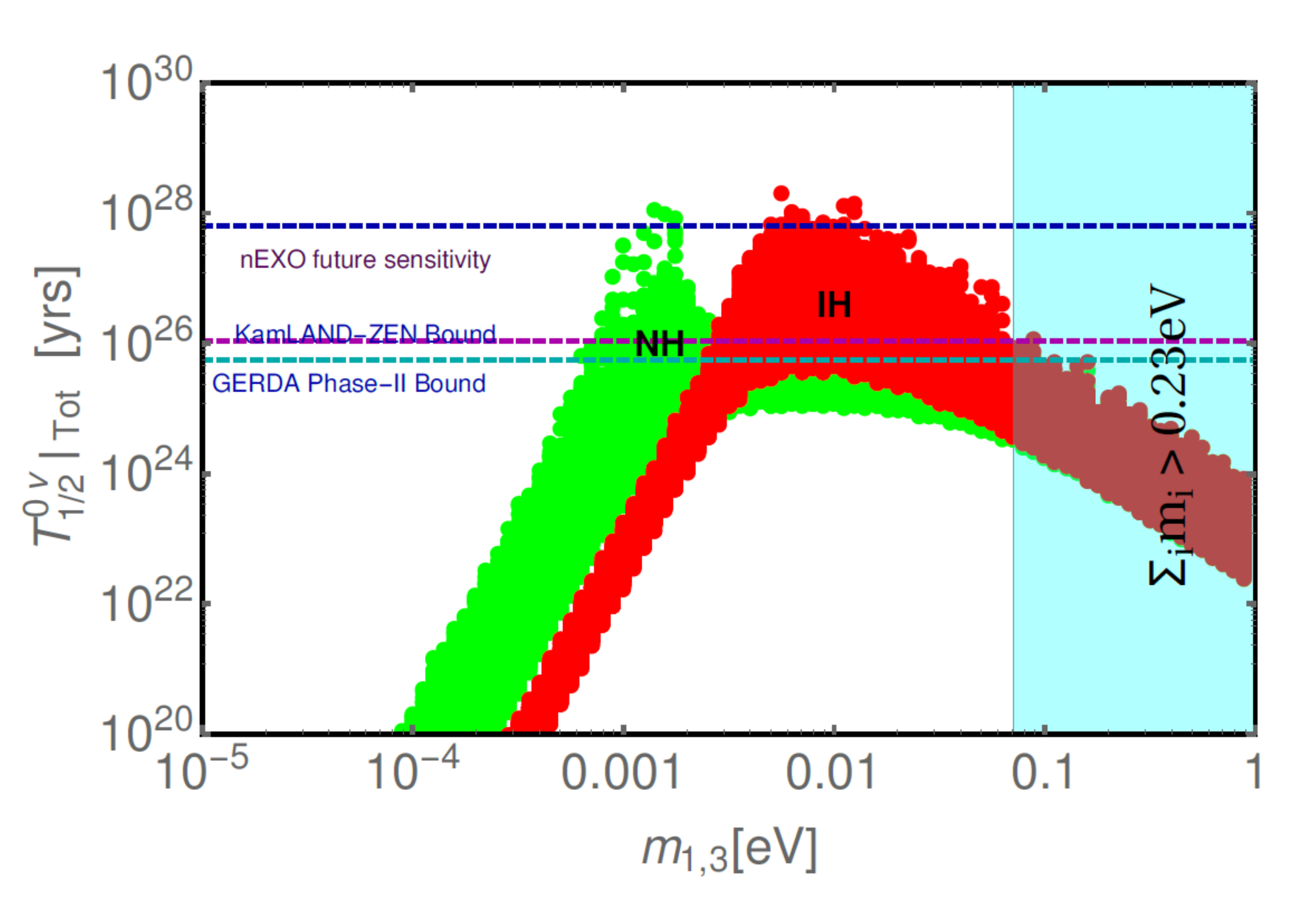}
\caption{Effective Majorana mass parameter and Half-life of $0\nu\beta\beta$ decay 
         due to combine effect of light active left-handed neutrinos, 
         heavy right-handed neutrinos and sterile neutrinos 
         as a function of the lightest neutrino mass for a NH and IH pattern.}
\label{plot:0nubb-5}
\end{center}
\end{figure}

Analogously one can write the combined contribution to half-life for $0\nu\beta\beta$ decay due to exchange of light left-handed neutrinos, 
heavy right-handed neutrinos and sterile neutrinos is given by 
\begin{align}
 [T_{1/2}^{0\nu}]^{-1} \!&=\! G_{01} 
 \left(|\mathcal{M}_\nu \eta_{\nu}|^2 + |\mathcal{M}_N \left(\eta_N +  \eta_S\right)|^2\right)\, ,  \nonumber \\
 &= \! G_{01}  \left| \frac{\mathcal{M}_\nu} {m_e} \right|^2 \cdot \left| m^{\rm Tot}_{\rm ee} \right|^2 \, ,
\end{align}
where the total contributions to effective mass parameter is found to be
\begin{align}
\label{mee:S} 
\left| m^{\rm Tot}_{\rm ee} \right|^2 =  \left|  \sum^{3}_{i=1} U^2_{ei} m_i  \right|^2+  \left| \sum^{3}_{i=1} \frac{\Theta^2_{ei} }{M_{i}}  \langle p^2 \rangle \right|^2 +
	\left|  \sum^{3}_{i=1} \frac{Y^2_{ei} }{M_{S_i}}  \langle p^2 \rangle \right|^2 \,.
\end{align}
Using these input parameters and the $3 \sigma$ ranges of the oscillation data from Table-\ref{table-osc},   we show  the variation of effective mass (left panel) and half-life (right panel) vs. lightest neutrino mass $m_{\rm lightest} = m_1~ ({\rm NH}),~ m_3$ (IH) in Fig.-\ref{plot:0nubb-1} due to the exchange of light Majorana neutrino. The green (red) region represents the contributions from light neutrinos obeying  Normal (Inverted) mass ordering. The cyan region is excluded due to the  95$\%$ CL limit $\sum_i m_i < 0.23 $ eV obtained from Planck+WMAP low multipole polarization+high resolution CMB+BAO data. The horizontal  bands/lines on the left (right) plot represent the upper (lower) limit on $m_{ee}$ ($T_{1/2}^{0 \nu}$) from various experiments. From these figures it can be noted that if only the light Majorana neutrinos contribute to the $o\nu \beta \beta$ decay process, it is very difficult to see the signal even at next generation experiments. On the other hand if the new physics contributions arising from the exchange of right-handed or sterile neutrinos are included, the $m_{ee}$ and $T_{1/2}^{0 \nu}$ values are significantly enhanced as seen from Figs. \ref{plot:0nubb-2} and \ref{plot:0nubb-3}.  The combined effects can saturate the current experimental limits as shown is Figs. \ref{plot:0nubb-4} and \ref{plot:0nubb-5}, which in turn give the lower bound on the lightest neutrino mass as $m_1 (m_3) \geq 1 ~(5)$ meV for NH (IH).

\section{Conclusion}
In this work, we have investigated gauge coupling unification within a non-supersymmetric  $SO(10)$ GUT with Pati-Salam symmetry and TeV scale asymmetric left-right theory as 
subgroups of the model. We have studied the RG evolution of gauge couplings for three different cases where the spontaneous symmetry breaking of asymmetric left-right 
symmetry is implemented with either Higgs doublets or triplets or combination of both. We found that, in the first case where  the intermediate symmetry  
$U(1)_R \times U(1)_{B-L} \to U(1)_Y$ is broken via the Higgs doublet ($H_R^0)$ around TeV scale, the unification of gauge couplings occurs at the scale $M_U \sim 10^{16}$ GeV. 
The Pati-Salam symmetry breaks at ${\cal O}(10^8)$ GeV, around which the right handed gauge boson $W_R$ gets its mass. The light neutrino masses are predominantly 
Dirac type and their sub-eV scale can be achieved  by suitable adjustment of different Yukawa couplings or by fine tuning of the parameters. Since there is no Majorana masses 
for the neutrinos, there will be no lepton number violation in this model. In the second case, the intermediate $U(1)_R \times U(1)_{B-L} \to U(1)_Y$ symmetry breaking is achieved 
through the Higgs triplets ($\Delta_{L,R}$), the unification occurs at $\sim 10^{16}$ GeV and the D-parity breaking scale is around 
$\sim 10^{12}$ GeV. Because of the lower D-parity breaking scale there is  sub-dominant Type-II seesaw contribution to the light neutrino mass along with the dominant type-II 
seesaw contribution with TeV scale right-handed neutrinos. Also, the $SU(2)_R$ breaking occurs at a lower scale ($\sim 10^5$ GeV), which allows right handed currents.
The inclusion of Higgs doublets and triplets for the intermediate $U(1)_R \times U(1)_{B-L} \to U(1)_Y$ symmetry breaking yields  almost similar results. With the asymmetric left-right symmetry 
breaking done with both combination of doublets and triplets, we have introduced one extra triplet to lower down the scale of D-parity breaking scale around $10^{10}-10^{11}~$GeV so that 
the type-II seesaw contribution to light neutrino mass i.e, $m^{II}_\nu= f v_L \approx \mathcal{O}(1) \cdot v^2 \cdot v_R / (M^\prime \cdot M_{D_P})$ is dominant.

We have also carried out a detail analysis on neutrino mass and neutrinoless double beta decay for the case where TeV scale asymmetric left-right symmetry breaking is done with 
both scalar doublets and triplets while adding an extra fermion singlet per generation to the minimal particle content. The resulting light neutrino mass formula is governed by 
natural type-II seesaw mechanism where mass eigenvalues for light and heavy neutrinos are related. The type-I seesaw contribution is exactly canceled out in the diagonalization 
of complete neutral lepton mass matrices and thus, there is no constraint on Dirac neutrino mass from light neutrino mass formula. We have considered Dirac neutrino mass matrix 
is equal to up-type quark mass matrix for all our numerical analysis which is a characteristics of Pati-Salam symmetry. It is well-known that  $0 \nu \beta \beta$ process violate lepton 
number  by two units and are mediated by Majorana neutrino mass terms which eventually manifest the Majorana nature of neutrinos. We found that if only the standard light active 
neutrinos are considered than the effective electron mass parameter $m_{ee}$ is few order  magnitude smaller that the current limit of the next generation experiments and it will be 
very difficult to get a signal.  Also we have shown that inclusion of  the new physics contributions to neutrinoless double beta decay induced by the exchange of right-handed
neutrinos and sterile neutrinos can saturate the present experimental limit and it is possible to see a signal at the next generation experiments. We also found the lower limits 
on the  absolute mass scale of the lightest neutrinos $m_1 \geq 1$ meV for NH and $m_3 \geq 5$ meV for IH.

\section{Acknowledgement}
SM  would like to thank University Grants Commission for financial support.  RM acknowledges the  support from the  Science and Engineering Research Board (SERB),
Government of India through grant No. SB/S2/HEP-017/2013. 
SP would like to acknowledge the warm hospitality provided by University 
of Hyderabad, India, between $22^{nd} - 29^{th}$ March,  2017, during which the part 
of this work is completed.

\bibliographystyle{utcaps_mod}
\bibliography{so10,so10a,so10b}
\end{document}